\documentclass[a4paper,12pt]{iopart}

\newcommand{\up}[1][]{_{\uparrow #1}}
\newcommand{\down}[1][]{_{\downarrow #1}}
\newcommand{\LD}{\epsilon}
\newcommand{\SP}{\xi}

\renewcommand{\vec}[1]{\mathbf{#1}}

\newif\ifPDF
\ifx\pdfoutput\undefined\PDFfalse
\else\ifnum\pdfoutput > 0 \PDFtrue
\else\PDFfalse
\fi
\fi

\ifPDF
\usepackage[pdftex]{graphicx,color}
\usepackage[pdftex]{hyperref}
\else
\usepackage[dvips]{graphicx}
\fi
\begin{document}

\title{Fermion pairing with spin-density imbalance in an optical lattice}
\date{\today}
\author{T. Koponen$^1$}
\author{J. Kinnunen$^1$}
\author{J.-P. Martikainen$^{2}$}
\author{L. M. Jensen$^1$}
\author{P. T\"{o}rm\"{a}$^1$}
\address{$^1$ Nanoscience Center, Department of Physics, P.O. Box 35, FI-40014 University
  of Jyv\"{a}skyl\"{a}, Finland}
\address{$^2$ Department of Physical Sciences, 
P.O. Box 64, FI-00014 University of Helsinki,  Finland}
\begin{abstract}
We consider pairing in a two-component atomic Fermi gas, in a three-dimensional optical lattice, when the components
have unequal densities, i.e.\ the gas is polarized. We show that a superfluid where the translational symmetry is
broken by a finite Cooper pair momentum, namely an FFLO-type state, minimizes the Helmholtz free energy of the system. We demonstrate that
such a state is clearly visible in the observable momentum distribution of the atoms, and analyze the dependence of the order parameter and the
momentum distribution on the filling fraction and the interaction strength.
\end{abstract}
\submitto{\NJP}
\maketitle

\section{Introduction}
\label{sec:intro}
Newly realized strongly interacting Fermi 
gases~\cite{Jochim2003b,Greiner2003a,Regal2004a,Zwierlein2004a,Bartenstein2004b,Kinast2004a,Chin2004a,Kinast2005a,Zwierlein2005a}
have enabled an intriguing line of research into the nature of fermion
superfluidity. In ultra-cold degenerate gases the system is composed
of atoms in different internal states. The number of atoms in
different states can be experimentally controlled and this has enabled
first experimental studies of polarized trapped Fermi
gases~\cite{Zwierlein2006a,Partridge2006a,Zwierlein2006b}. Imbalanced
atomic gases have recently inspired a number of theoretical works
\cite{Sheehy2006a,Pieri2006a,Kinnunen2006a,Yi2006a,Chevy2006a,He2006a,DeSilva2006a,Haque2006a,Ho2006a,Bulgac2006a,Machida2006a,Jensen2006a,DeSilva2006b,Imambekov2006a},
in addition to the wide literature on fermion pairing with unequal
chemical potentials or number densities in the fields of condensed
matter, nuclear and high-energy physics (for a review of some of these
studies, see \cite{Casalbuoni2004a}). 

While several parameters of a harmonically trapped Fermi gas can be
controlled experimentally, the inhomogeneity is often crucial
and some parameters, such as the atomic mass, cannot be changed.
The use of optical lattices provides unprecedented tunability and, for example,
the effective masses of atoms can be changed by simply changing the laser intensities.
Furthermore, in the center of the optical lattice the effects due to the 
harmonic trapping can be weak. This versatile tool has been recently widely used
and has enabled, among other things, the observation 
of the superfluid-Mott insulator quantum phase transition~\cite{Greiner2002a}
in a cloud a bosons. Use of optical lattices is not restricted by quantum
statistics and indeed fermions have been studied 
experimentally in one-dimensional~\cite{Modugno2003a,Pezze2004a} as well as in
three-dimensional~\cite{Kohl2005a,Stoferle2006a,Stoferle2006b}
optical lattices. In addition, the first experimental results of Bose-Fermi mixtures
in optical lattices were recently reported~\cite{Gunter2006a}.

In this paper we investigate the properties of polarized Fermi gases
in a three dimensional optical lattice. In particular, we focus
on the states of constant density and compare in detail the energetics
of the breached pair (BP) states~\cite{Sarma1963a,Liu2003a,Liu2004a,Forbes2005a}
and the simplest variant of the FFLO
states~\cite{Fulde1964a,Larkin1965b} in a three-dimensional optical
lattice. We find that, while the non-zero gap BP state can be a
minimum of the Helmholtz free energy it can lower its energy by
forming pairs with non-zero momentum.  

We show that the lattice dispersion causes a different dependence of
the superfluid gap on the polarization compared to homogeneous
systems.  We find that, while the non-zero gap BP state can be a minimum of the
Helmholtz free energy it can lower its energy by breaking the
translational symmetry and forming pairs with non-zero momentum as in
FFLO states. Section 2. reviews the theory used in the numerical
calculations. In section 3. we discuss the free energy analysis and
explain, illustrated by numerical examples, why it is necessary to use
the Helmholtz free energy in case of a system isolated from the
environment with respect to particle exchange, as is the case with
trapped atoms. In section 4. we present the results on the
FFLO states: we analyze the system behaviour as a function of the
lattice filling fraction and the interaction strength between the two pairing
components. Furthermore, we present the momentum distribution of the
atoms, showing that this easily observable quantity carries a clear signature of
the FFLO state. We conclude by a discussion in section 5.

\section{Theory}
\label{sec:theory}
The microscopic Hamiltonian for fermions of two (pseudo) spins $\uparrow$
and $\downarrow$, in an external potential
\(V(\vec{x})\), is
\begin{equation}
  \label{eq:microscopic_hamiltonian}
\fl  H = \sum_{\sigma}\int
  \psi_{\sigma}^\dagger(\vec{x})\left(-\frac{\hbar^2}{2m}\nabla^2 +
    V(\vec{x}) \right)\psi_{\sigma}(\vec{x})\,d^3\vec{x} +
  4\pi\hbar^2\frac{a}{m}\int
  \psi\up^\dagger(\vec{x})\psi\down^\dagger(\vec{x})\psi\down(\vec{x})\psi\up(\vec{x})\,d^3\vec{x}, 
\end{equation}
where \(a\) is the $s$-wave scattering length. With a sufficiently deep periodic
potential, the atoms are localized in the minima of the potential, and
the system can be described by the Fermi-Hubbard Hamiltonian \cite{Jaksch1998a}, 
\begin{eqnarray}
\fl \nonumber H -\mu\up N\up - \mu\down N\down =& -\sum_\vec{n} \left(\mu\up\hat{c}\up[\vec{n}]^\dagger \hat{c}\up[\vec{n}] +
\mu\down\hat{c}\down[\vec{n}]^\dagger \hat{c}\down[\vec{n}] \right)+ U\sum_\vec{n} \hat{c}\up[\vec{n}]^\dagger \hat{c}\down[\vec{n}]^\dagger
\hat{c}\down[\vec{n}]\hat{c}\up[\vec{n}]
\\ &-\left(J_x \sum_{\langle \vec{n},\vec{m}\rangle_x}  + J_y \sum_{\langle
    \vec{n},\vec{m}\rangle_y} + J_z \sum_{\langle \vec{n},\vec{m}\rangle_z}\right)
\left(\hat{c}\up[\vec{m}]^\dagger \hat{c}\up[\vec{n}] +
  \hat{c}\down[\vec{m}]^\dagger \hat{c}\down[\vec{n}] \right).
\end{eqnarray}
Here \(\langle \vec{n},\vec{m}\rangle_x\) means a nearest neighbour
pair in the \(x\)-direction.
In mean-field theory the interaction term in the Hamiltonian is approximated with
\begin{equation}
\fl U\sum_\vec{n} \hat{c}\up[\vec{n}]^\dagger \hat{c}\down[\vec{n}]^\dagger
   \hat{c}\down[\vec{n}]\hat{c}\up[\vec{n}] = U\sum_\vec{n} \left(\left\langle\hat{c}\up[\vec{n}]^\dagger
   \hat{c}\down[\vec{n}]^\dagger\right\rangle \hat{c}\down[\vec{n}]\hat{c}\up[\vec{n}] + \hat{c}\up[\vec{n}]^\dagger \hat{c}\down[\vec{n}]^\dagger
   \left\langle\hat{c}\down[\vec{n}]\hat{c}\up[\vec{n}]\right\rangle - \left\langle\hat{c}\up[\vec{n}]^\dagger
   \hat{c}\down[\vec{n}]^\dagger\right\rangle\left\langle\hat{c}\down[\vec{n}]\hat{c}\up[\vec{n}]\right\rangle
 \right),
\end{equation}
where the Hartree and Fock terms have been dropped since the former
are effectively included in the chemical potentials and the latter do not contribute.
A general ansatz \(U
\left\langle\hat{c}\down[\vec{n}]\hat{c}\up[\vec{n}]\right\rangle =
\Delta e^{2i\vec{q}\cdot\vec{n}}\) describes several apparently
different states. For equal (pseudo)spin number densities, \(\vec{q} = 0\) gives the
standard Bardeen-Cooper-Schrieffer (BCS) solution. For unequal densities \(\vec{q} = 0\) gives the
Breached Pairing (BP) solution
and non-zero
values for \(\vec{q}\) describe a Fulde-Ferrel-Larkin-Ovchinnikov (FFLO)
state where Cooper pairs have a finite momentum, \(\vec{q}\). Finally,
\(\Delta = 0\) naturally describes a state with no superfluid,
i.e. the normal state.

Including the chemical potentials in the Hamiltonian it becomes, in momentum
space,
\begin{equation}
  \label{eq:hamiltonian}
\fl H =\frac{1}{M} \sum_{\vec{k}}\left(\left(\LD_{\vec{k}} - \mu\up \right)\hat{c}\up[\vec{k}]^\dagger \hat{c}\up[\vec{k}] +
 \left(\LD_{\vec{k}} - \mu\down
 \right)\hat{c}\down[\vec{k}]^\dagger \hat{c}\down[\vec{k}] +
\Delta \hat{c}\up[\vec{q}+\vec{k}]^\dagger
\hat{c}\down[\vec{q}-\vec{k}]^\dagger + h.c.
 -\frac{|\Delta|^2}{U} \right) ,
\end{equation}
where \(M\) is the number of lattice sites,
\(\mu\up\) and \(\mu\down\) are the chemical potentials of the
different spin species and the lattice dispersion is given by 
\begin{equation}
\LD_{\vec{k}} = 2J_x(1-\cos(k_x d))+2J_y(1-\cos(k_y d)) + 2J_z(1-\cos(k_z d)).
\end{equation}
Here $d$ is the lattice parameter, i.e. the distance between two
neighbouring lattice points.
The parameter \(U\) describes the energy associated with the
interaction of the particles, with negative values corresponding to an
attractive interaction. The hopping parameter, \(J\) is the energy
gain corresponding to tunneling between nearest neighbour sites. In a
lattice, \(J\) is essentially the band width.
For a more detailed discussion on the parameters \(J\) and \(U\), see
e.g. \cite{Koponen2006a}. Using the fermionic anticommutators, the Hamiltonian
can be rearranged as (note periodic boundary conditions in the $k$ summations)
\begin{eqnarray}
  \label{eq:hamiltonian2}
\nonumber H =\frac{1}{M} \sum_{\vec{k}}\Bigg(&\left(\LD_{\vec{k}+\vec{q}} - \mu\up \right)\hat{c}\up[\vec{k}+\vec{q}]^\dagger \hat{c}\up[\vec{k}+\vec{q}] 
 -\left(\LD_{-\vec{k}+\vec{q}} - \mu\down
 \right)\left(\hat{c}\down[-\vec{k}+\vec{q}]\hat{c}\down[-\vec{k}+\vec{q}]^\dagger - 1\right)  +\\
&\Delta \hat{c}\up[\vec{q}+\vec{k}]^\dagger
\hat{c}\down[\vec{q}-\vec{k}]^\dagger + \Delta^* \hat{c}\down[\vec{q}-\vec{k}]\hat{c}\up[\vec{q}+\vec{k}]
 -\frac{|\Delta|^2}{U} \Bigg).
\end{eqnarray}
This can be written in matrix form as 
\begin{eqnarray}
  \label{eq:hamiltonianmatrix}
  \nonumber H =&\frac{1}{M} \sum_{\vec{k}}\left(\begin{array}{cc} \hat{c}\up[\vec{k}+\vec{q}]^\dagger &
      \hat{c}\down[-\vec{k}+\vec{q}] \end{array}\right)
  \left(\begin{array}{cc}\LD_{\vec{k}+\vec{q}} - \mu\up & \Delta
      \\ \Delta^* &  -\left(\LD_{-\vec{k}+\vec{q}} - \mu\down
      \right) \end{array}
  \right)\left(\begin{array}{c}\hat{c}\up[\vec{q}+\vec{k}]\\\hat{c}\down[\vec{q}-\vec{k}]^\dagger \end{array} \right)\\ 
&+ \frac{1}{M}\sum_{\vec{k}} \left(\LD_{-\vec{k}+\vec{q}} - \mu\down - \frac{|\Delta|^2}{U}\right).
\end{eqnarray}
Because $\Delta$ was chosen as the amplitude of the order parameter, it is a real
number, which simplifies the
expressions. The second sum in \eref{eq:hamiltonianmatrix} is just a
constant, but it is important for the calculation of free
energies. The eigenvalues of the matrix part (i.e. the quasiparticle energies) are
\begin{equation}
  \label{eq:eigenvalues}
  E_{\vec{k},\vec{q},\pm} = \frac{\mu\down - \mu\up}{2} +
  \frac{\LD_{\vec{k}+\vec{q}}-\LD_{-\vec{k}+\vec{q}}}{2} \pm
  \sqrt{\left(
      \frac{\LD_{\vec{k}+\vec{q}}+\LD_{-\vec{k}+\vec{q}}}{2}
      - \frac{\mu\down + \mu\up}{2}\right)^2 + \Delta^2}.
\end{equation}
With the introduction of the single particle energies \(\SP\up[\vec{k}] =
\LD_{\vec{k}} - \mu\up\) and \(\SP\down[\vec{k}] =
\LD_{\vec{k}} -\mu\down\), this can be written in a simpler form
\begin{equation}
  \label{eq:eigenvaluessimpler}
  E_{\vec{k},\vec{q},\pm} = \frac{\SP\up[\vec{k}+\vec{q}] -
    \SP\down[-\vec{k}+\vec{q}]}{2} \pm \sqrt{ \left(\frac{\SP\up[\vec{k}+\vec{q}] +
    \SP\down[-\vec{k}+\vec{q}]}{2}\right)^2 + \Delta^2}.
\end{equation}
Unequal densities introduce unequal
chemical potentials, which destroys the particle-hole symmetry that
exists in the standard BCS theory. A suitable Bogoliubov
transformation,
\begin{equation}
  \label{eq:bogoliubov}
  \left(\begin{array}{c}\hat{c}\up[\vec{q}+\vec{k}]\\\hat{c}\down[\vec{q}-\vec{k}]^\dagger \end{array} \right) = 
\left(\begin{array}{cc}
    u_{\vec{k},\vec{q}} & v_{\vec{k},\vec{q}} \\ -v_{\vec{k},\vec{q}} & u_{\vec{k},\vec{q}}
  \end{array}\right)
\left(\begin{array}{c}\hat{\gamma}\up[\vec{k},\vec{q}] \\ \hat{\gamma}\down[-\vec{k},\vec{q}]^\dagger\end{array}\right),
\end{equation}
with the coefficients given by
\begin{eqnarray}
  \label{eq:bogoliubovcoefficients}
  \nonumber u_{\vec{k},\vec{q}}^2 &= \frac{1}{2}\left( 1 +
    \frac{\xi_A}{\sqrt{\xi_A^2 + \Delta^2}}\right)\\
  \nonumber v_{\vec{k},\vec{q}}^2 &= \frac{1}{2}\left( 1 -
    \frac{\xi_A}{\sqrt{\xi_A^2 + \Delta^2}}\right)\\
   u_{\vec{k},\vec{q}}v_{\vec{k},\vec{q}} &= \frac{\Delta}{2\sqrt{\xi_A^2 + \Delta^2}},
\end{eqnarray}
where \(\xi_A = (\SP\up[\vec{k}+\vec{q}]+\SP\down[-\vec{k}+\vec{q}])/2\),
 diagonalizes the Hamiltonian to
\begin{equation}
  \label{eq:diagonalhamiltonian}
\fl  H =\frac{1}{M} \sum_{\vec{k}}\left[\left(\begin{array}{cc}
      \hat{\gamma}\up[\vec{k},\vec{q}]^\dagger &
      \hat{\gamma}\down[-\vec{k},\vec{q}]\end{array}\right)\left(\begin{array}{cc} E_{\vec{k},\vec{q},+} & 0 \\ 0 & E_{\vec{k},\vec{q},-} 
\end{array}\right)
  \left(\begin{array}{c}
      \hat{\gamma}\up[\vec{k},\vec{q}] \\
      \hat{\gamma}\down[-\vec{k},\vec{q}]^\dagger\end{array}\right) + \left(\SP\down[-\vec{k}+\vec{q}] - \frac{\Delta^2}{U}\right)\right].
\end{equation}
Here the \(\hat{\gamma}_{\sigma,\vec{k},\vec{q}}\) are the quasiparticle
operators and they obey the Fermi statistics.

\subsection{Self-consistent equations for $\Delta$, $\mu\up$, and
  $\mu\down$}
The ansatz \(U
\left\langle\hat{c}\down[\vec{n}]\hat{c}\up[\vec{n}]\right\rangle =
\Delta e^{2i\vec{q}\cdot\vec{n}}\) implies \(\Delta = U\sum_{\vec{k}}\left\langle
  \hat{c}\down[-\vec{k}+\vec{q}]
  \hat{c}\up[\vec{k}+\vec{q}]\right\rangle/M\), where \(M\) is the
total number of lattice sites. With the Bogoliubov
transformation \eref{eq:bogoliubov}, this can be written as
\begin{eqnarray}
  \label{eq:gap_equation}
\nonumber  \Delta &= \frac{U}{M}\sum_{\vec{k}}\left\langle
  \hat{c}\down[-\vec{k}+\vec{q}]
  \hat{c}\up[\vec{k}+\vec{q}]\right\rangle = \frac{U}{M}\sum_{\vec{k}}
u_{\vec{k},\vec{q}}v_{\vec{k},\vec{q}}\left(f(E_{\vec{k},\vec{q},-}) -
  f(E_{\vec{k},\vec{q},+}) \right)\\
&=  \Delta\frac{U}{M}\sum_{\vec{k}}\frac{f(E_{\vec{k},\vec{q},-}) -
  f(E_{\vec{k},\vec{q},+})}{2\sqrt{\xi_A^2 + \Delta^2}}.
\end{eqnarray}
The following equations hold for the particle numbers:
\begin{eqnarray}
  \label{eq:numberup}
  N\up &= \sum_{\vec{k}}\left\langle \hat{c}\up[\vec{k}]^\dagger
    \hat{c}\up[\vec{k}]\right\rangle =  \sum_{\vec{k}}
  u_{\vec{k},\vec{q}}^2 f(E_{\vec{k},\vec{q},+}) + v_{\vec{k},\vec{q}}^2
  f(E_{\vec{k},\vec{q},-})\\
\label{eq:numberdown}
  N\down &= \sum_{\vec{k}}\left\langle \hat{c}\down[\vec{k}]^\dagger
    \hat{c}\down[\vec{k}]\right\rangle = \sum_{\vec{k}}
  u_{\vec{k},\vec{q}}^2 f(-E_{\vec{k},\vec{q},-}) + v_{\vec{k},\vec{q}}^2
  f(-E_{\vec{k},\vec{q},+}).
\end{eqnarray}
Equations \eref{eq:gap_equation}, \eref{eq:numberup}, and
\eref{eq:numberdown} are \emph{self-consistent} equations in the sense
that they are always satisfied by the stable macroscopic
state. Especially the gap equation, \eref{eq:gap_equation}, can have
many solutions, of which the one with the lowest free energy is stable. For
example, for equal populations of the two components, even below the
critical temperature, the gap equation always has the trivial \(\Delta
= 0\) solution, in addition to the standard BCS solution. However, the
BCS solution has lower free energy and is therefore stable. 

It is useful to define a gap function \(g(\Delta)\) that vanishes at the
correct value of \(\Delta\). This reduces \eref{eq:gap_equation} to \(g(\Delta) = 0\).
\begin{equation}
  \label{eq:gap_function}
  g(\Delta) = \Delta\frac{U}{M}\sum_{\vec{k}}\frac{f(E_{\vec{k},\vec{q},-}) -
  f(E_{\vec{k},\vec{q},+})}{2\sqrt{\xi_A^2 + \Delta^2}} - \Delta = 0.
\end{equation}

In optical lattices, the relevant quantity, instead of the total
number of particles in one component, is the filling fraction of one
component. The filling fraction is defined as the number of particles
divided by the number of lattice sites. We denote filling fractions as
\(f\up = N\up /M\) and \(f\down = N\down /M\). Because of the Pauli exclusion principle,
only one fermion of each kind can fit on the lowest Hubbard band at
each site, thereby having both \(f\up\) and \(f\down\) equal to \(1\) means
having a full lattice, which is a band insulator. Putting three atoms
in the same site or on the same momentum state would mean populating
higher energy bands, which involves a significant cost in energy. For
superfluidity, the optimum setting, giving the highest value for
\(\Delta\), is \(f\up = 0.5\) and \(f\down = 0.5\)\cite{Micnas1990a}. 
For population imbalance, it is useful to define the polarization:
\begin{equation}
P = \frac{f\up - f\down}{f\up + f\down}.
\end{equation}

\subsection{Experimental parameters}
In our numerical calculations we consider $^6$Li atoms trapped in a
lattice created by \(\lambda = 1030\)~nm lasers. The wavelength is
related to the lattice parameter $d$ by $d = \lambda/2$. The lattice height
is \(2.5 E_R\), where the recoil energy is
\(\hbar^2(2\pi/\lambda)^2/2m\). In the present paper calculations are
done in zero temperature. Except where explicitly noted, we use
\(-1000\) Bohr radii for the scattering length \(a\). Typical
lattice sizes we have used are \(64\times64\times64\) and \(128\times 128
\times 128\). We calculate the finite lattice sums explicitly, which
limits all the vectors in our numerical analysis to a set of discrete
lattice points.
 
\section{Free energy analysis}
\label{sec:free_energy}
The relevant free energy of a Fermi gas depends on the physical system
in question. We start by considering the
grand potential \(\Omega = -\frac{1}{\beta}\ln Z_G\), where \(Z_G =
\Tr e^{-\beta H}\), is the partition function of the grand canonical
ensemble. The
partition function is \cite{FetterWalecka}
\begin{equation}
  \label{eq:Z}
  Z_G = \Tr e^{-\beta H} = \prod_{\vec{k}} \left(1+ e^{-\beta
      \frac{E_{\vec{k},\vec{q},+}}{M}} \right)\left(1+ e^{\beta
      \frac{E_{\vec{k},\vec{q},-}}{M}} \right)e^{-\beta C},
\end{equation}
where \(C\) corresponds to the constant part of the Hamiltonian and
\(\beta = 1/k_{\mathrm{B}} T\). The grand potential is
\begin{equation}
  \label{eq:grand_potential}
\fl  \Omega = \frac{1}{M}\sum_{\vec{k}}\left[ \SP\down[-\vec{k}+\vec{q}] +
  E_{\vec{k},\vec{q},-} -
    \frac{\Delta^2}{U}\right] - \frac{1}{\beta} \sum_{\vec{k}}\left[\ln\left(1 + e^{-\beta
      \frac{E_{\vec{k},\vec{q},+}}{M}} \right) + \ln\left(1 + e^{\beta
      \frac{E_{\vec{k},\vec{q},-}}{M}} \right)\right]. 
\end{equation}
At low temperatures this is independent of \(\beta\) and becomes
\begin{equation}
  \label{eq:grand_potential_zero_T}
\fl  \Omega = \frac{1}{M}\sum_{\vec{k}}\left( \SP\down[-\vec{k}+\vec{q}] +
  E_{\vec{k},\vec{q},-} - \frac{\Delta^2}{U}
  + E_{\vec{k},\vec{q},+}\Theta(-E_{\vec{k},\vec{q},+})
  - E_{\vec{k},\vec{q},-}\Theta(E_{\vec{k},\vec{q},-})\right).
\end{equation}

There are two different schemes for treating population imbalanced
Fermi gases: fixing the particle numbers or fixing the chemical
potentials; this choice depends on the physical system in question. 
We are interested in the former, but first we will briefly
discuss the latter, to enlighten the differences between the two cases.
\subsection{Fixed chemical potentials, $\vec{q} = 0$}
\begin{figure}
\begin{center}
\includegraphics{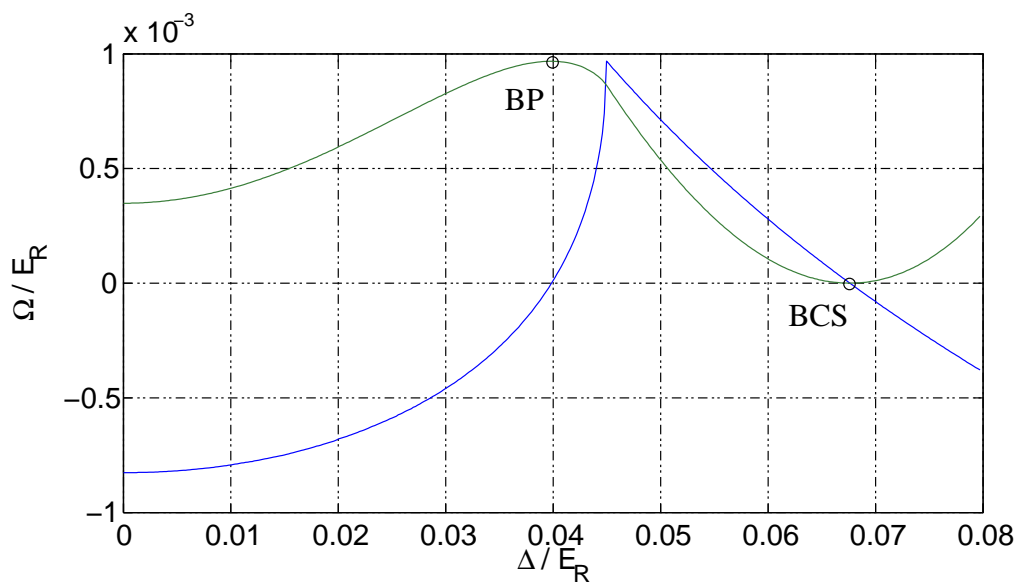}
\caption{The grand potential $\Omega$ per lattice site (green), in the units
  of recoil energy, and $g(\Delta)/\Delta$ (blue)
  in arbitrary units. The maximum of $\Omega$ at $\Delta = 0.04$ corresponds to the
  BP state with polarization $0.12$, and the minimum at $\Delta = 0.068$ corresponds to the BCS
  state with a zero polarization. The chemical potentials are fixed at
  $\mu\up = 0.394$ and $\mu\down = 0.304$.}
\label{fig:grand_potential}
\end{center}
\end{figure}

\begin{figure}
\begin{center}
\includegraphics[width=0.7\textwidth]{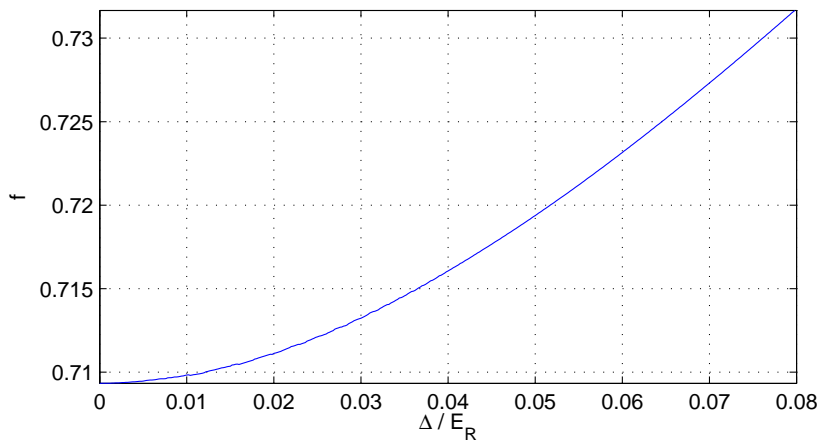}
\caption{The total filling fraction $f = f\up + f\down$ as a function
  of $\Delta$, with the chemical potentials fixed at $\mu\up = 0.394$ and $\mu\down = 0.304$. }
\label{fig:total_number}
\end{center}
\end{figure}

With fixed chemical potentials, the relevant thermodynamic free energy
is the grand potential \(\Omega(\Delta,\mu\up,\mu\down)\). The extrema of \(\Omega\),
$\partial \Omega/\partial \Delta = 0$, 
correspond to solutions of the gap equation,
\eref{eq:gap_function}, i.e. $g(\Delta) = 0$. Figure \ref{fig:grand_potential} shows a
typical scenario with the chemical potentials fixed at \(\mu\up =
0.394\) and \(\mu\down = 0.304\). The figure shows how the extrema of
\(\Omega\) coincide with the zeros of \(g(\Delta)\). In this scheme,
the BP state is unstable because it is a local maximum of
\(\Omega(\Delta)\). However, both the polarization and the total
number of atoms change as \(\Delta\) changes, as can be seen from
figure \ref{fig:total_number}. Therefore the extrema correspond to
situations with different total numbers of particles. If the numbers of
atoms are to stay fixed in the system, this comparison is not valid.

\subsection{Fixed numbers of particles, $\vec{q}=0$}
When numbers of particles, instead of the chemical potentials, are
fixed, the relevant thermodynamical quantity is the Helmholtz free energy,
\begin{equation}
  \label{eq:helmholz}
  F(\Delta,N\up,N\down) = \Omega + \mu\up \frac{N\up}{M} + \mu\down
  \frac{N\down}{M} = \Omega + \mu\up f\up + \mu\down f\down.
\end{equation}
Now the physical solutions are the minima of
\(F(\Delta,N\up,N\down)\): $\partial F/\partial \Delta = 0$ gives the extrema and $\partial^2 F/\partial \Delta^2
> 0$ defines the minima. These
solutions again coincide with the zeros of \(g(\Delta)\). Moreover, the extrema given by $F$ are the same as given by $\Omega$, i.e.\ $\partial 
\Omega/\partial \Delta = \partial F/\partial \Delta = 0$, but
determining which of the extrema are minima, and thereby the physical solutions, depends on 
whether $F$ or $\Omega$ is used. The
solutions where \(\vec{q}\) is fixed at zero are the uniform
solutions. If the densities of the different components are the same,
the solution is known as the BCS state and if the densities differ,
the solution is known as the BP state. With fixed densities, both the
BCS ($N\up = N\down$) and the BP ($N\up \neq N\down$) state correspond to a minimum of
\(F(\Delta)\). We consider there the case $N_\uparrow \neq N_\downarrow$, that is, the BP state.

\begin{figure}
  \begin{center}
    \includegraphics{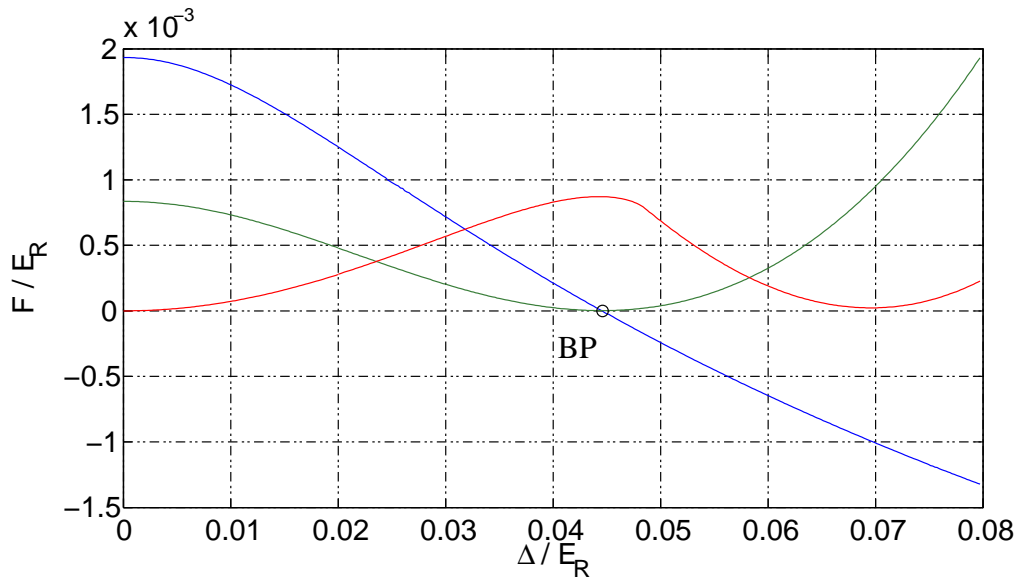}
    \caption{The Helmholtz free energy $F(\Delta)$ per lattice site
      (green), in the units of recoil energy, and $g(\Delta) / \Delta$
      (blue) in arbitrary units. The filling fractions are fixed at
      $f\up = 0.44$ and $f\down = 0.36$, so the polarization is
      $0.1$. The minimum of $F$ at $\Delta \approx 0.045$, coinciding with the zero of $g(\Delta)$, corresponds
      to the BP state. Also shown is $\Omega(\Delta)$ (red) with chemical
      potentials fixed so that the filling fractions have the values
      mentioned above at the point where \(g(\Delta) = 0\).}
  \label{fig:helmholz}
  \end{center}
\end{figure}

Figure \ref{fig:helmholz} shows the Helmholtz free energy and the gap
function with fixed numbers of particles (with $N\up \neq N\down$) and \(\vec{q}\) fixed to \(0\). It is clear that
\(F(\Delta)\) is minimized with a finite \(\Delta\), i.e.\ the BP state is stable in this consideration, however one has to consider also the 
possibility of a non-zero $\vec{q}$. 

\section{FFLO-states}
Stability analysis is always limited to some set of states. Many
states have been studied in isotropic
systems.\cite{Sarma1963a,Fulde1964a,Larkin1965b,Machida1984a,Bedaque2003a,Sedrakian2005a}
 A Monte Carlo
study \cite{Dukelsky2006a} suggests the FFLO state to be the ground
state in a two-dimensional lattice in the weakly interacting
regime. Here we study BCS, BP, and single mode FFLO states in
three-dimensional lattices.

\begin{figure}
  \centering
  \includegraphics[width=0.48\textwidth]{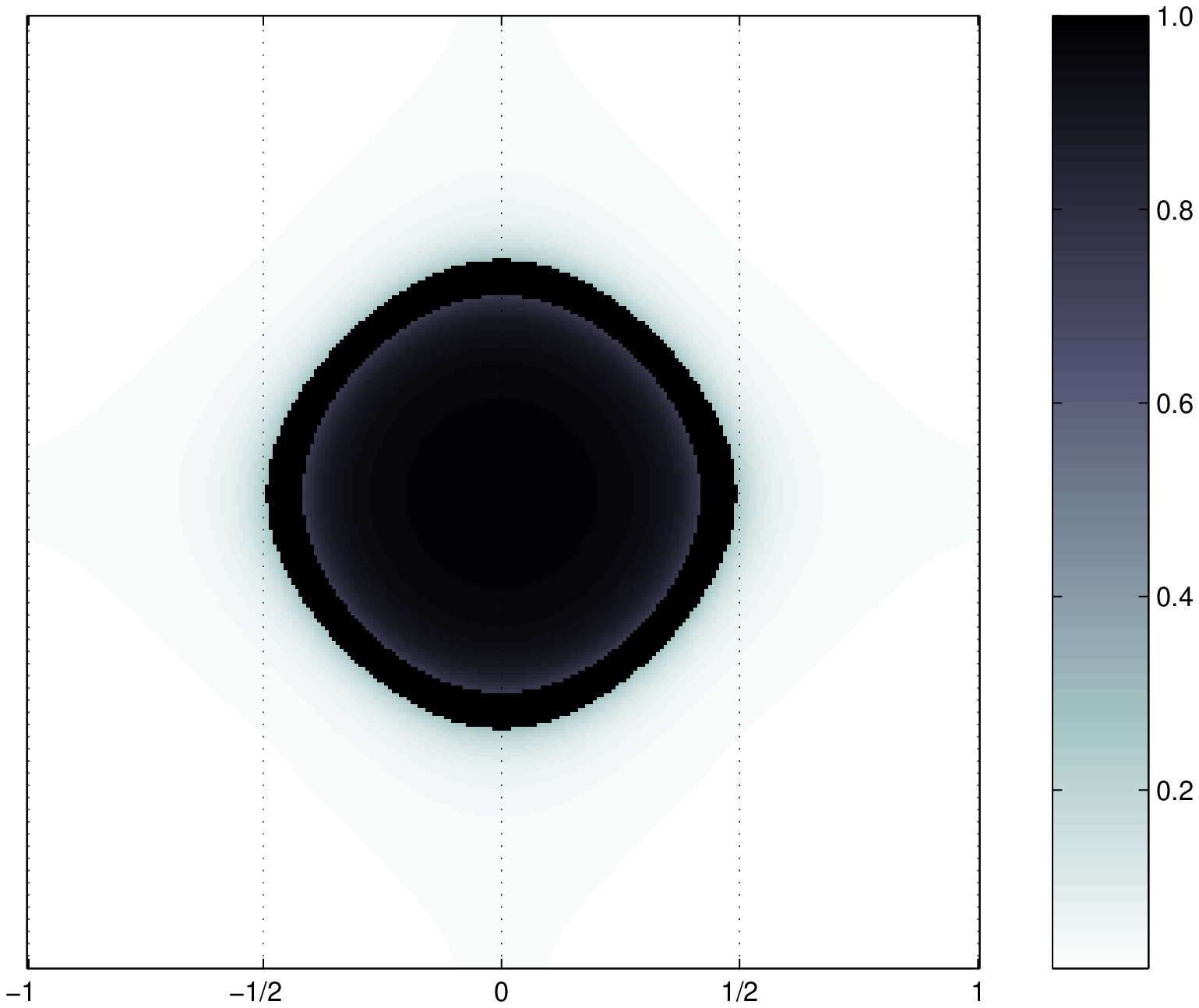}
  \includegraphics[width=0.48\textwidth]{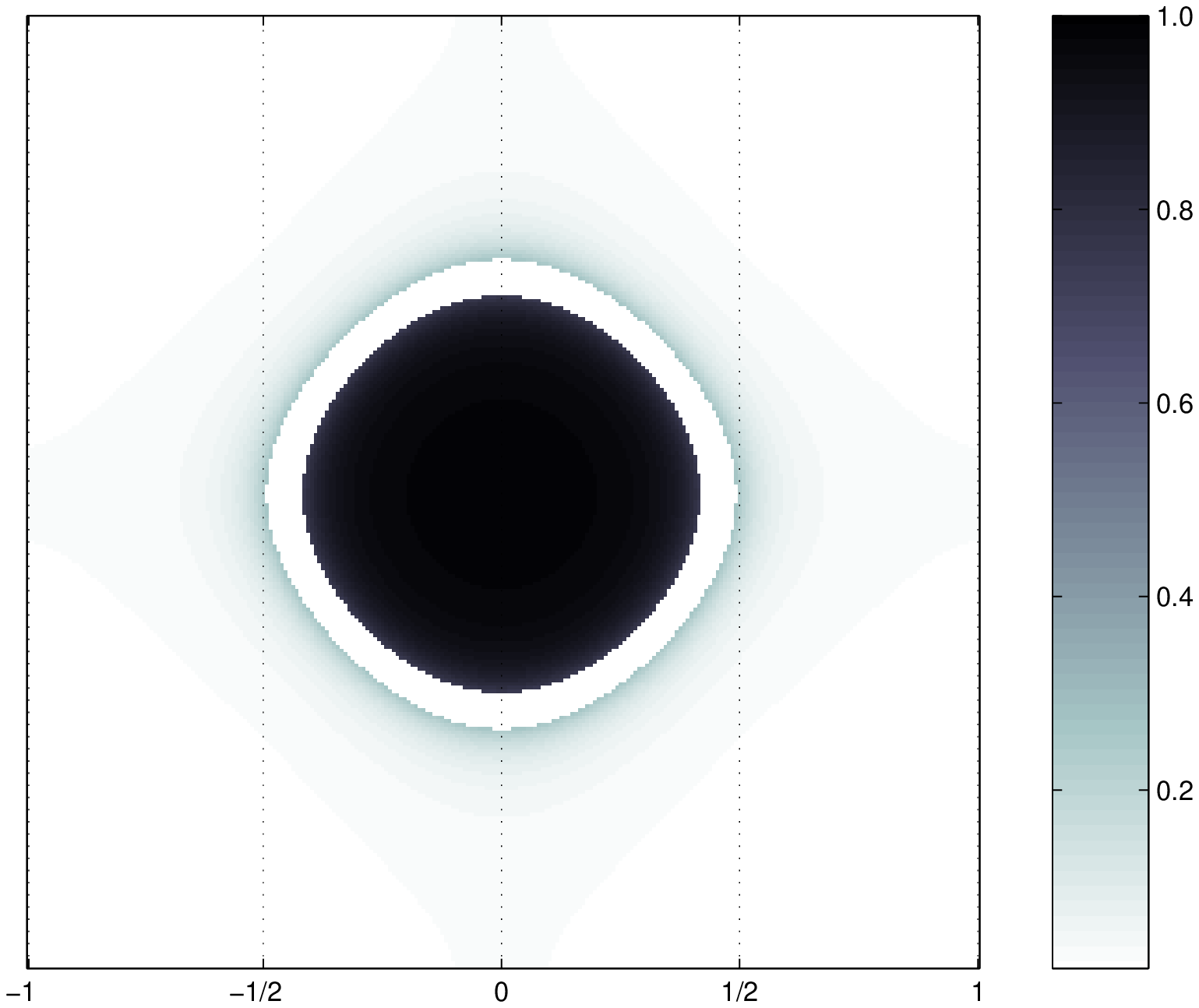}
  
  \includegraphics[width=0.48\textwidth]{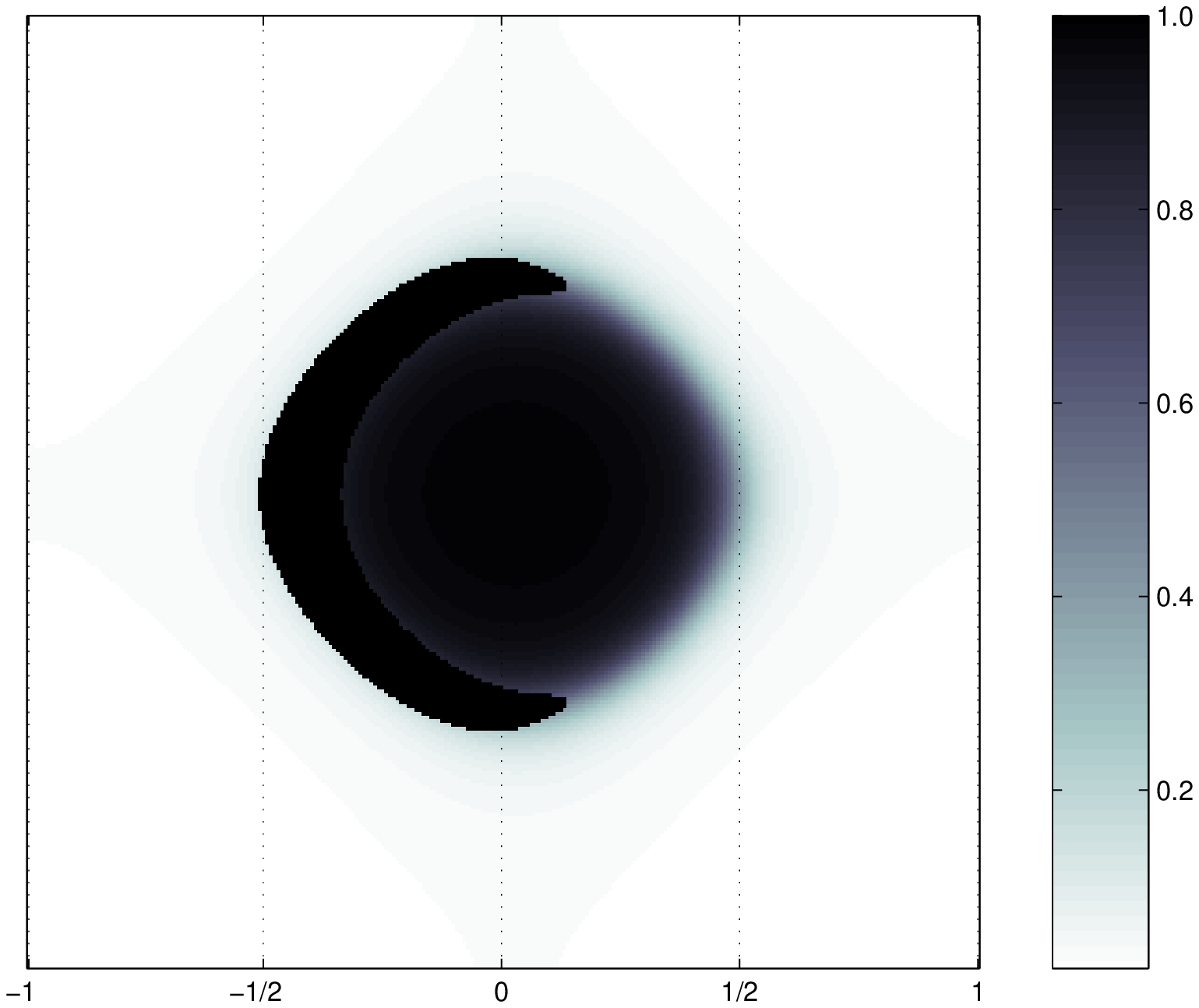}
  \includegraphics[width=0.48\textwidth]{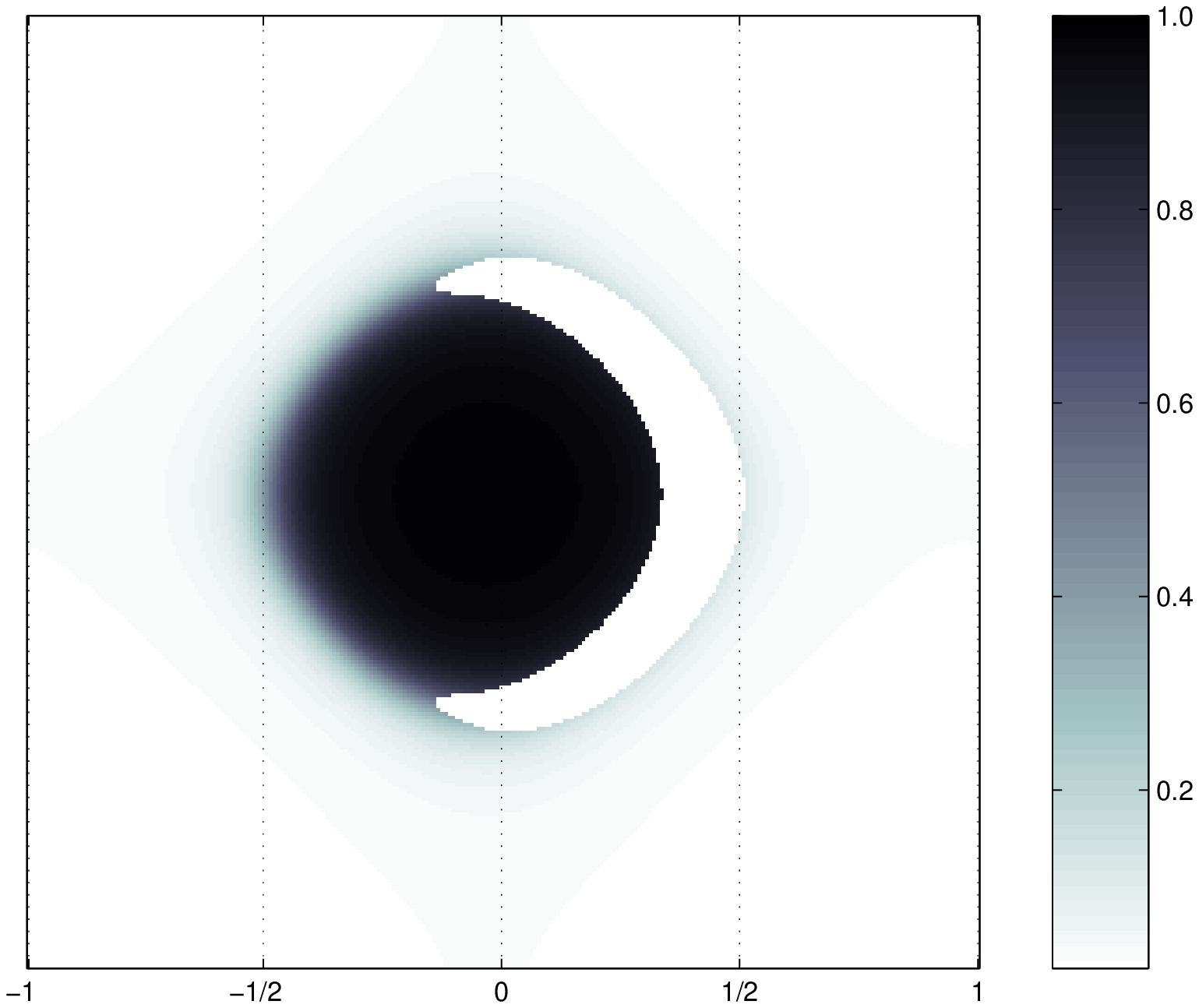}
  \caption{The momentum distributions $f\up$ and $f\down$, on the $k_z
    = 0$ plane, of the BP (top row) and FFLO (bottom row) states. Both
    states demonstrate a depairing region around the Fermi surface,
    but whereas the region is symmetric in the BP state, it is
    asymmetric in FFLO. The background scattering length is $-1500\
    a_0$, the total filling fraction is $0.1$ and polarization
    $0.2$. In this situation, the FFLO state is energetically favorable.} 
  \label{fig:breachedpairing}
\end{figure}

When the particle numbers are fixed and the momentum of the Cooper
pairs, $2\vec{q}$, is allowed to get non-zero values, the
translational symmetry is broken and the state is
FFLO-like. The stable state is now found by minimizing \(F\) with
respect to \(\Delta\) and \(\vec{q}\). Figure
\ref{fig:breachedpairing} shows the momentum distributions of the
different components along the $k_z = 0$ plane in BP and FFLO
states. Note that there is a vacant region, or breach, in the momentum
distribution of the minor component in the BP phase.

\begin{figure}
\begin{center}
\includegraphics{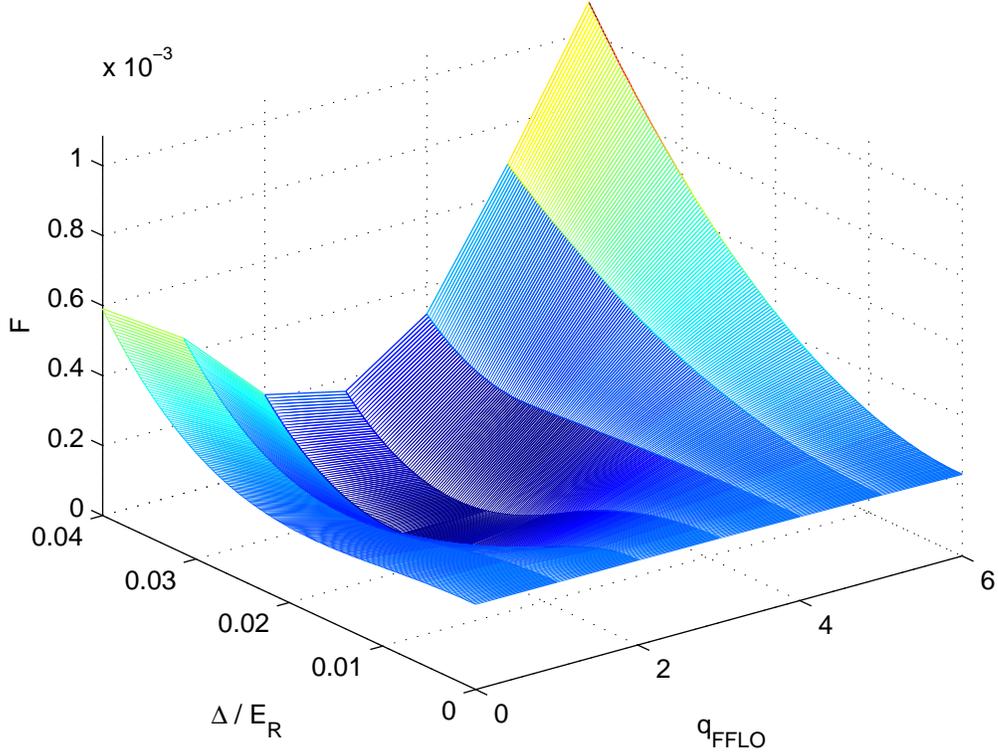}
\caption{The Helmholtz free energy $F(\Delta,q)$ per lattice
  site. Here $\vec{q}$ is in the $x$-direction, i.e. $\vec{q} = (q,0,0)$. The filling fractions are
  fixed at $f\up = 0.24$ and $f\down = 0.16$, so that $P = 0.2$. 
The units of $q$ are selected so that $(32,32,32)$ would correspond to
the corner of the first Brillouin zone, the R point.}
\label{fig:fflo_energy}
\end{center}
\end{figure}

Figure \ref{fig:fflo_energy} shows a typical free energy landscape for
the FFLO state. The filling fractions are fixed at \(f\up = 0.24\) and
\(f\down = 0.16\). The figure shows that the minimum energy is found
with a non-zero \(\Delta\) and non-zero \(\vec{q}\). The BP state is a
saddle point and the BCS solution is absent since it does not support
polarization in zero temperature.

\begin{figure}
  \centering
  \includegraphics[width=0.7\textwidth]{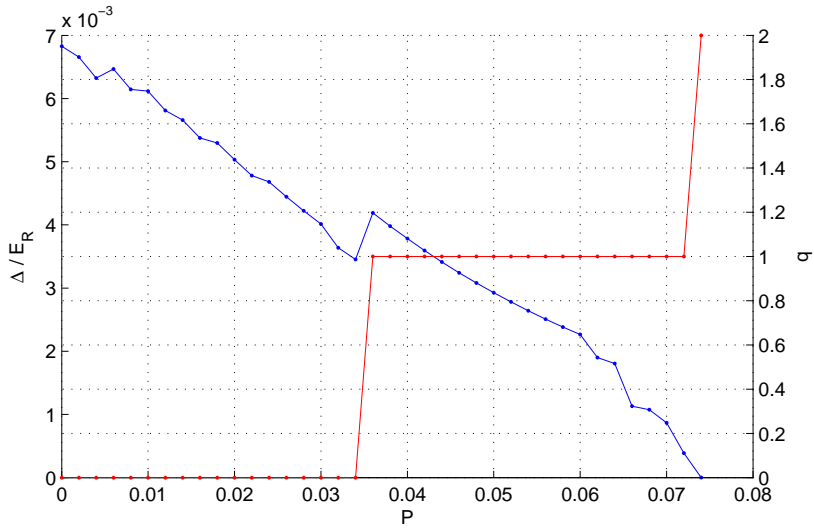}
  \caption{The order parameter $\Delta$ (blue) and the magnitude of
    \(\vec{q}\) (red) as a function of
    polarization $P$. Here the total filling fraction is $0.1$, the
    scattering length is $-1000 a_0$, and the lattice size is
    $128\times 128 \times 128$.
    The $q$ is given in reciprocal lattice indices, so that
    $64$ on the $y$-axis would correspond to the edge of the first
    Brillouin zone.}
  \label{fig:q-steps}
\end{figure}

\begin{figure}
\begin{center}
\includegraphics[width=0.7\textwidth]{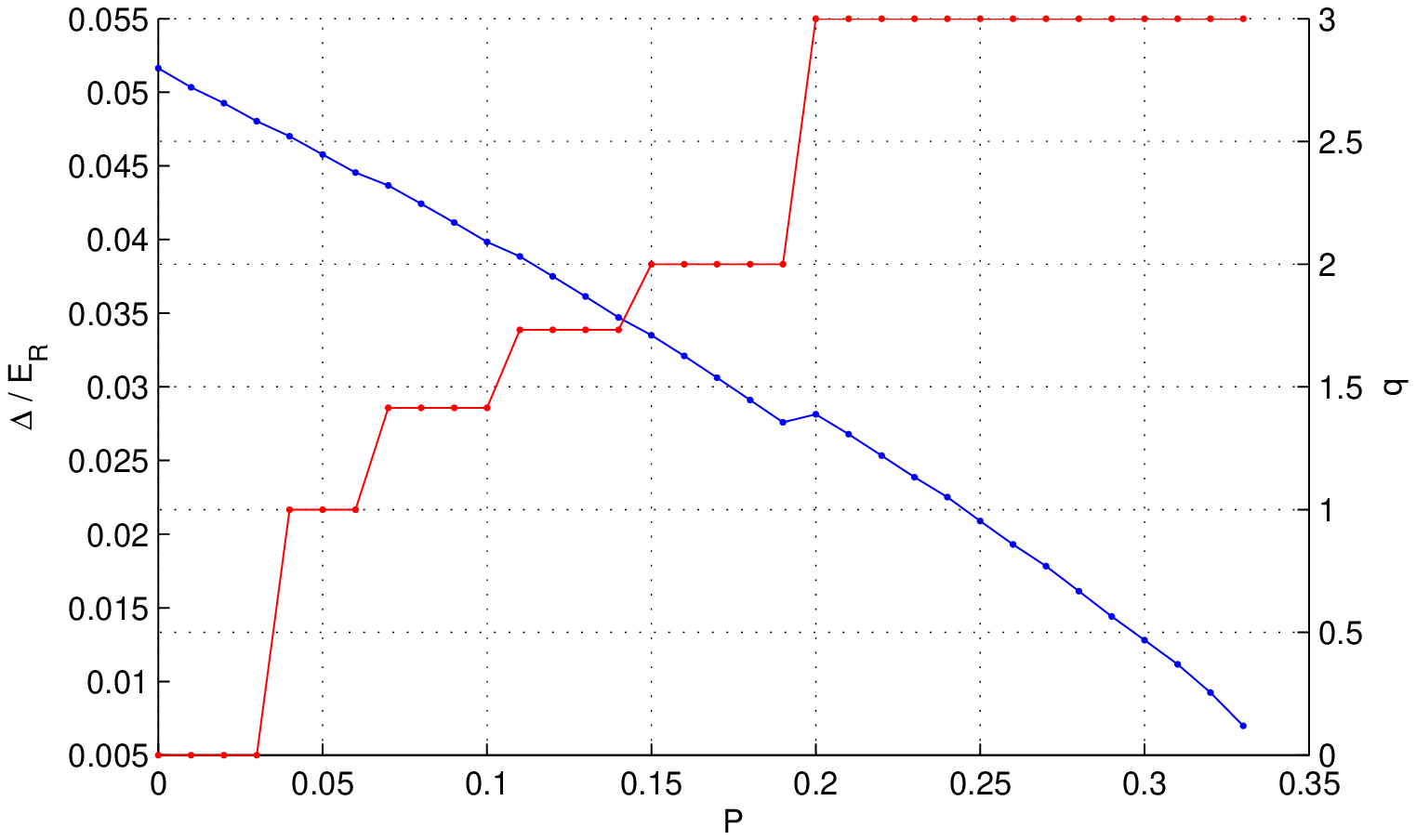}
\caption{The energy gap $\Delta$ and the magnitude of $\vec{q}$ as a function of polarization with
  the total filling fraction of $f = 0.4$ and scattering length $a = -1000
  a_0$.}
\label{fig:fflo_delta_a-1000f0.2}
\end{center}
\end{figure}

\begin{figure}
\begin{center}
\includegraphics[width=0.7\textwidth]{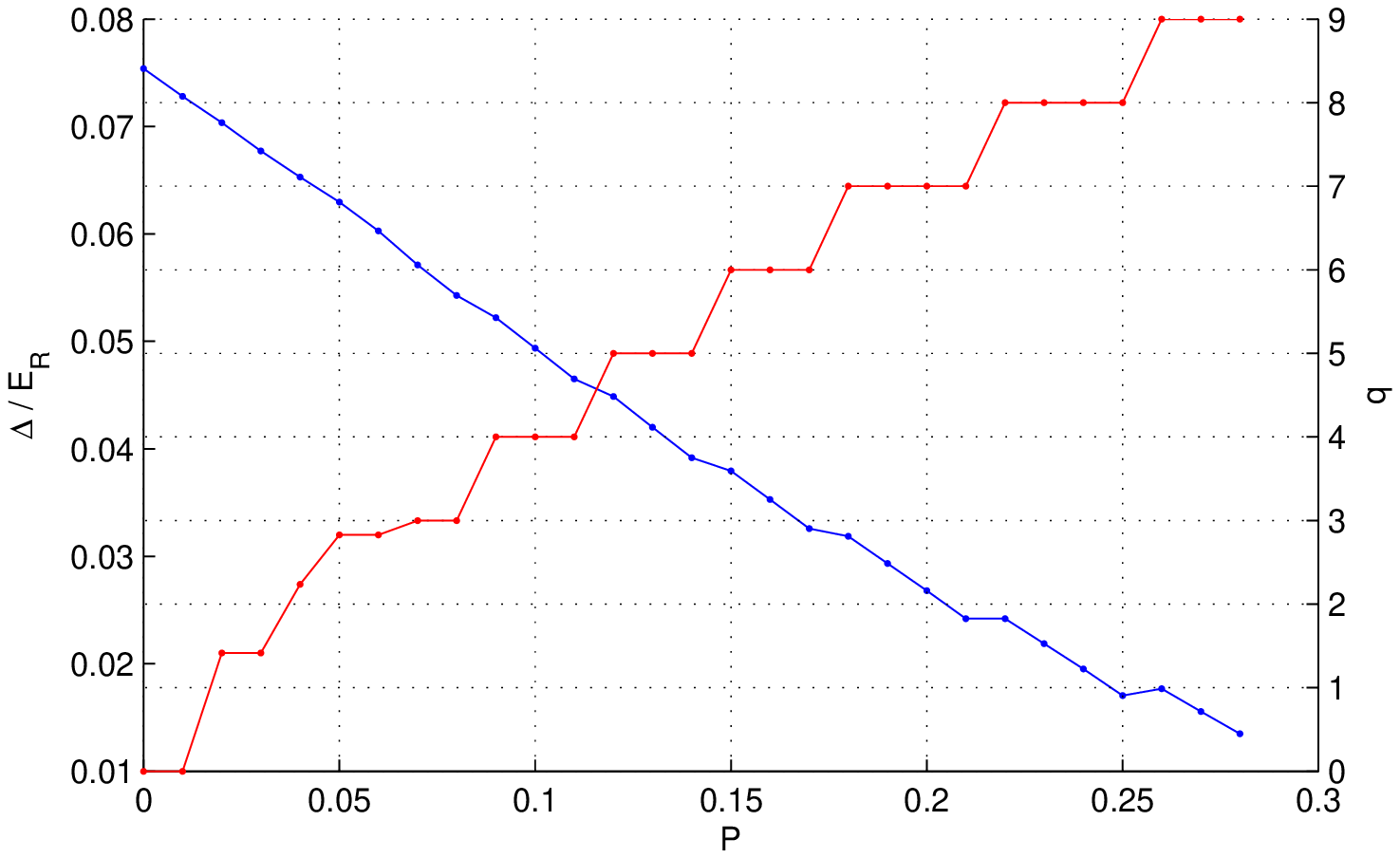}
\caption{The energy gap $\Delta$ and the magnitude of $\vec{q}$ as a function of polarization with
  the total filling fraction of $f = 1.0$ and scattering length $a = -1000
  a_0$.}
\label{fig:fflo_delta_a-1000f0.5}
\end{center}
\end{figure}

\begin{figure}
\begin{center}
\includegraphics[width=0.7\textwidth]{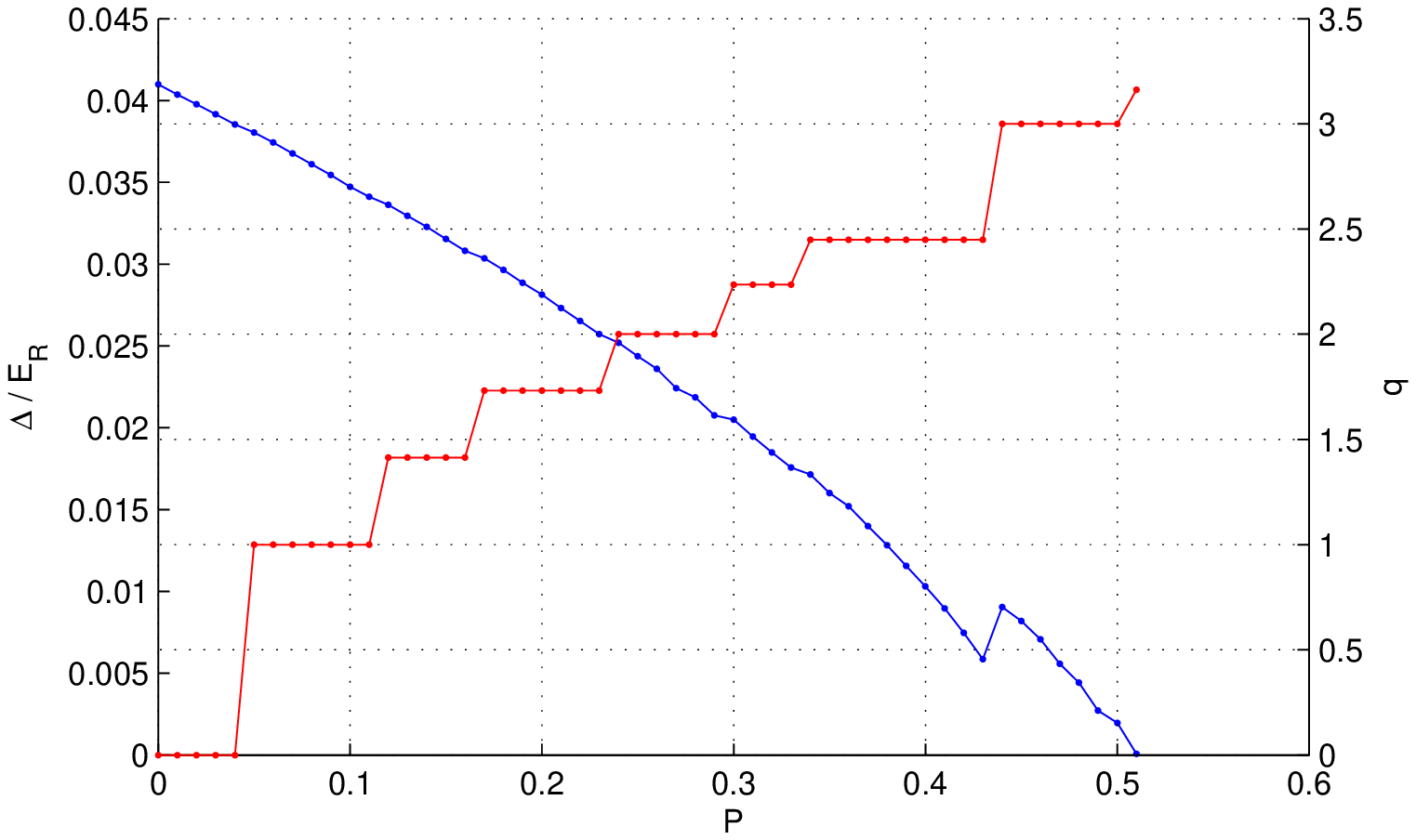}
\caption{The energy gap $\Delta$ and the magnitude of $\vec{q}$ as a function of polarization with
  the total filling fraction of $f = 0.1$ and scattering length $a = -1500
  a_0$.}
\label{fig:fflo_delta_a-1500f0.05}
\end{center}
\end{figure}

\begin{figure}
\begin{center}
\includegraphics[width=0.7\textwidth]{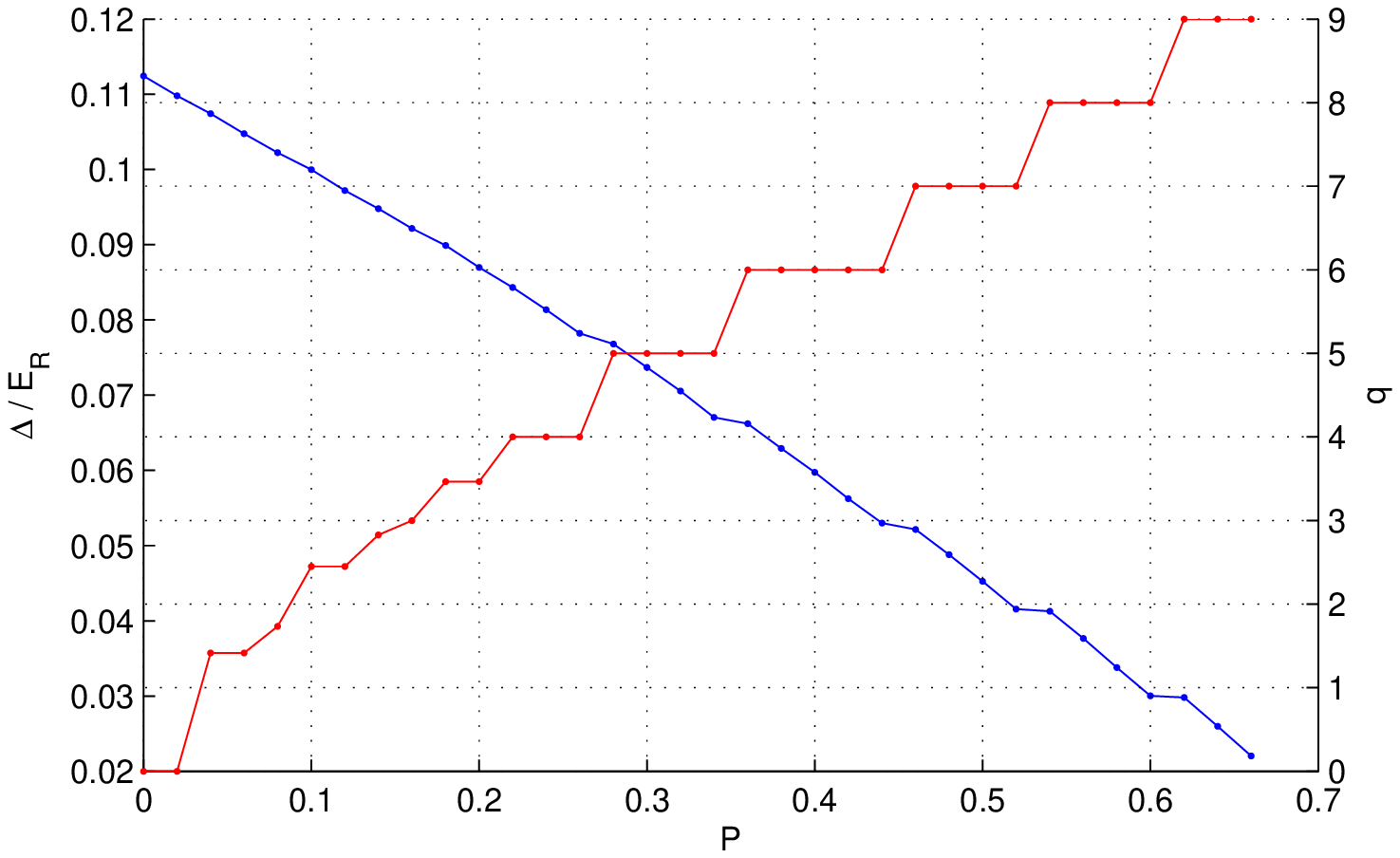}
\caption{The energy gap $\Delta$ and the magnitude of $\vec{q}$ as a function of polarization with
  the total filling fraction of $f = 0.4$ and scattering length $a = -1500
  a_0$.}
\label{fig:fflo_delta_a-1500f0.2}
\end{center}
\end{figure}

\begin{figure}
\begin{center}
\includegraphics[width=0.7\textwidth]{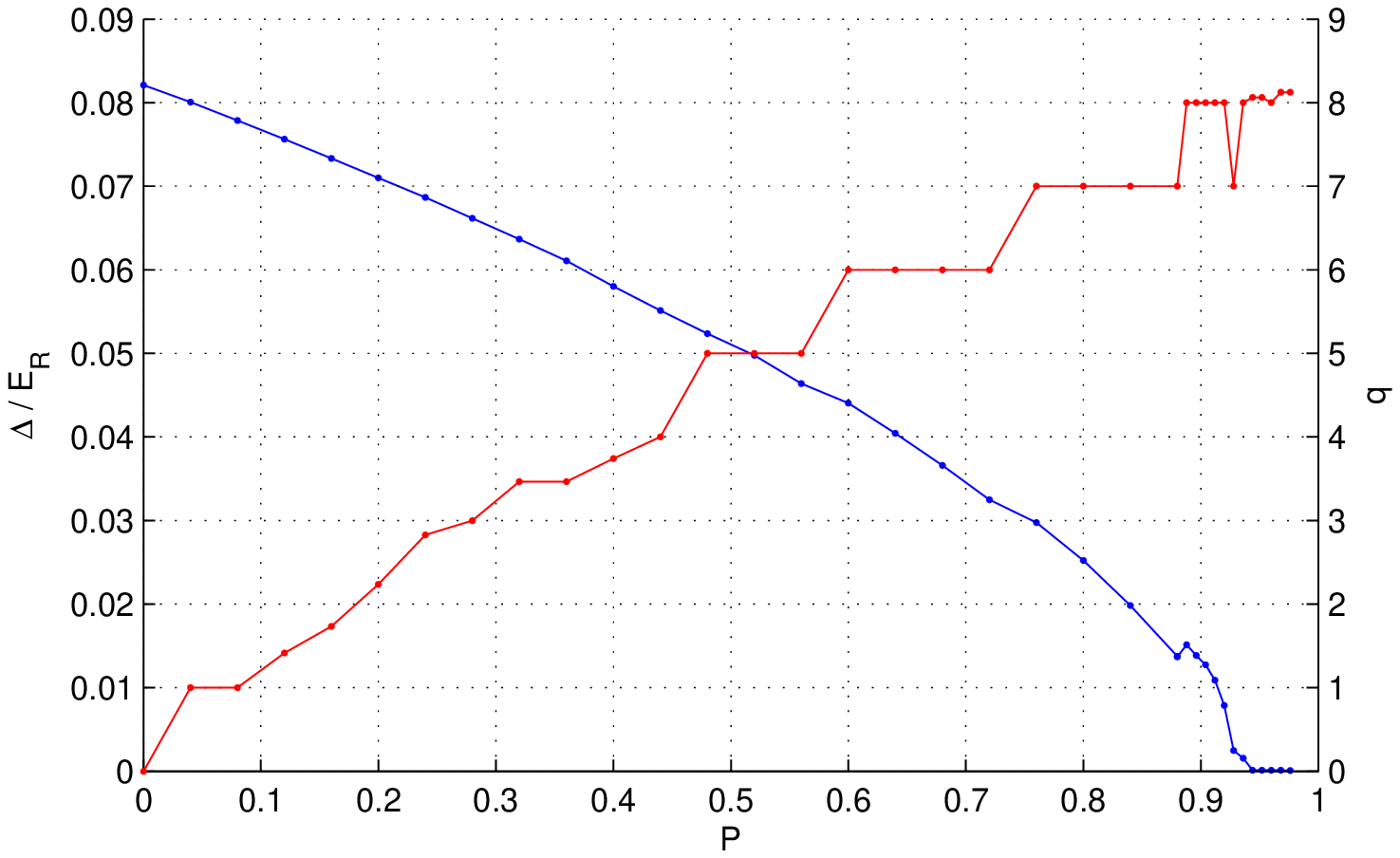}
\caption{The energy gap $\Delta$ and the magnitude of $\vec{q}$ as a function of polarization with
  the total filling fraction of $f = 0.1$ and scattering length $a = -2000
  a_0$.}
\label{fig:fflo_delta_a-2000f0.05}
\end{center}
\end{figure}

Figures \ref{fig:q-steps} -
\ref{fig:fflo_delta_a-2000f0.05} show the energy gap $\Delta$ and the
FFLO momentum $\vec{q}$ as a function of polarization for different
interaction strengths. With increasing polarization, the energy gap
$\Delta$ decreases and the magnitude of $\vec{q}$ increases. Whether a critical polarization $P_{\mathrm{c}}$,
where $\Delta$ vanishes, exists, depends on the total density $f\up
+ f\down$ and the interaction strength between the atoms, characterized by the scattering length $a$. With $a = -1000$ Bohr
radii and total filling fractions between \(0.4\) and \(1.0\), the
critical polarizations are around $0.3$. Raising the scattering
length to $a = -1500\ a_0$ gives a $P_{\mathrm{C}}$ around $0.6$
with the same densities. With $a = -2000\ a_0$, $P_{\mathrm{c}}$ is
more than  $0.9$, as can be seen in figure \ref{fig:fflo_delta_a-2000f0.05}.

Figures \ref{fig:q-steps} - \ref{fig:fflo_delta_a-1500f0.2} show the $\vec{q}=0$, i.e. the BP phase, with small
polarizations. However, the discrete steps in the values of $\vec{q}$
are due to the finite size of the lattice and using a larger lattice
allows for smaller steps with shorter intervals. We expect these steps, and
the kinks in the energy gap, to vanish in the limit of larger
lattices. The figures also show that stronger interactions as well as
larger densities lead to higher critical polarizations, as is
expected. 

\begin{figure}
\begin{center}
\includegraphics[width=0.48\textwidth]{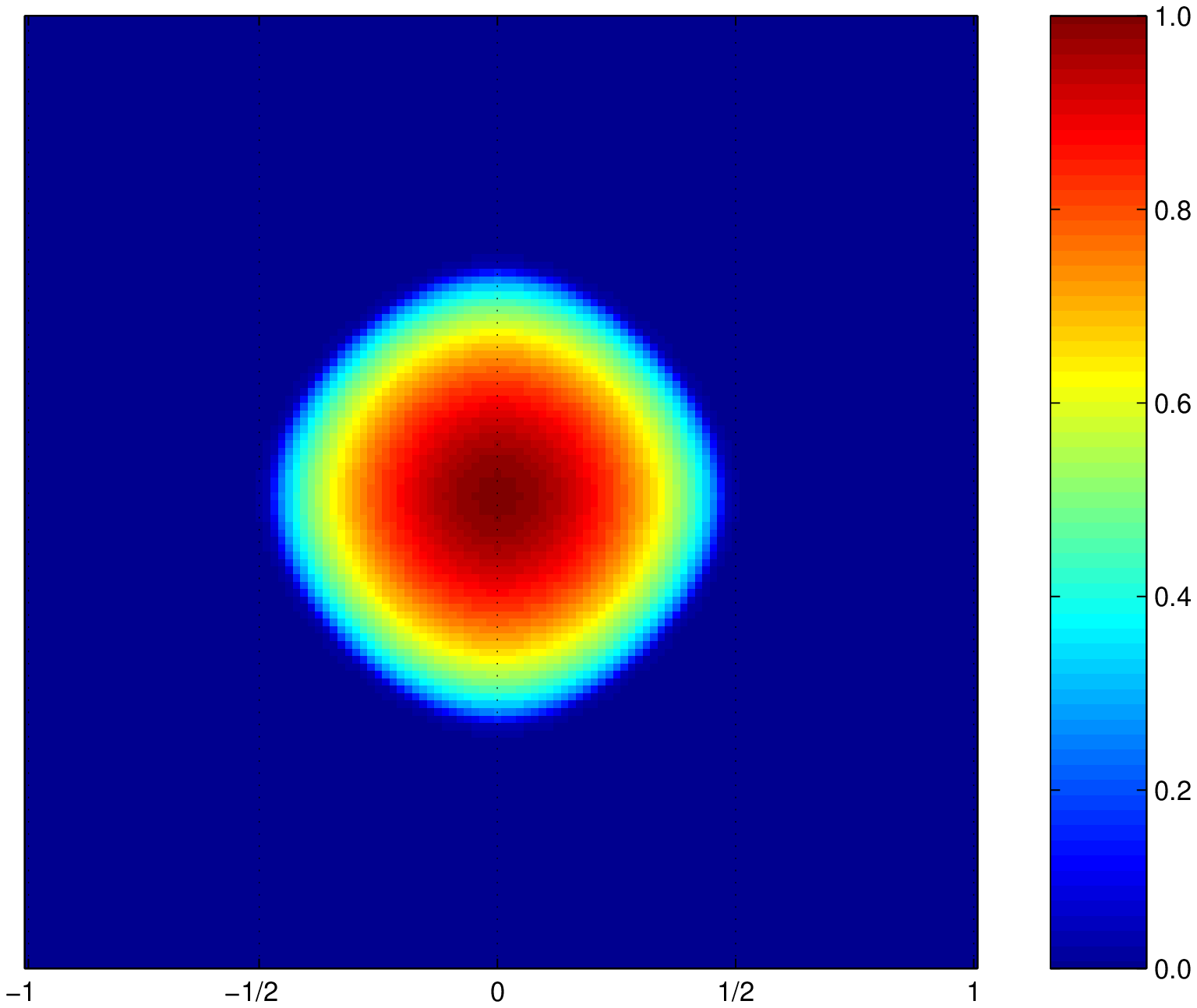}
\includegraphics[width=0.48\textwidth]{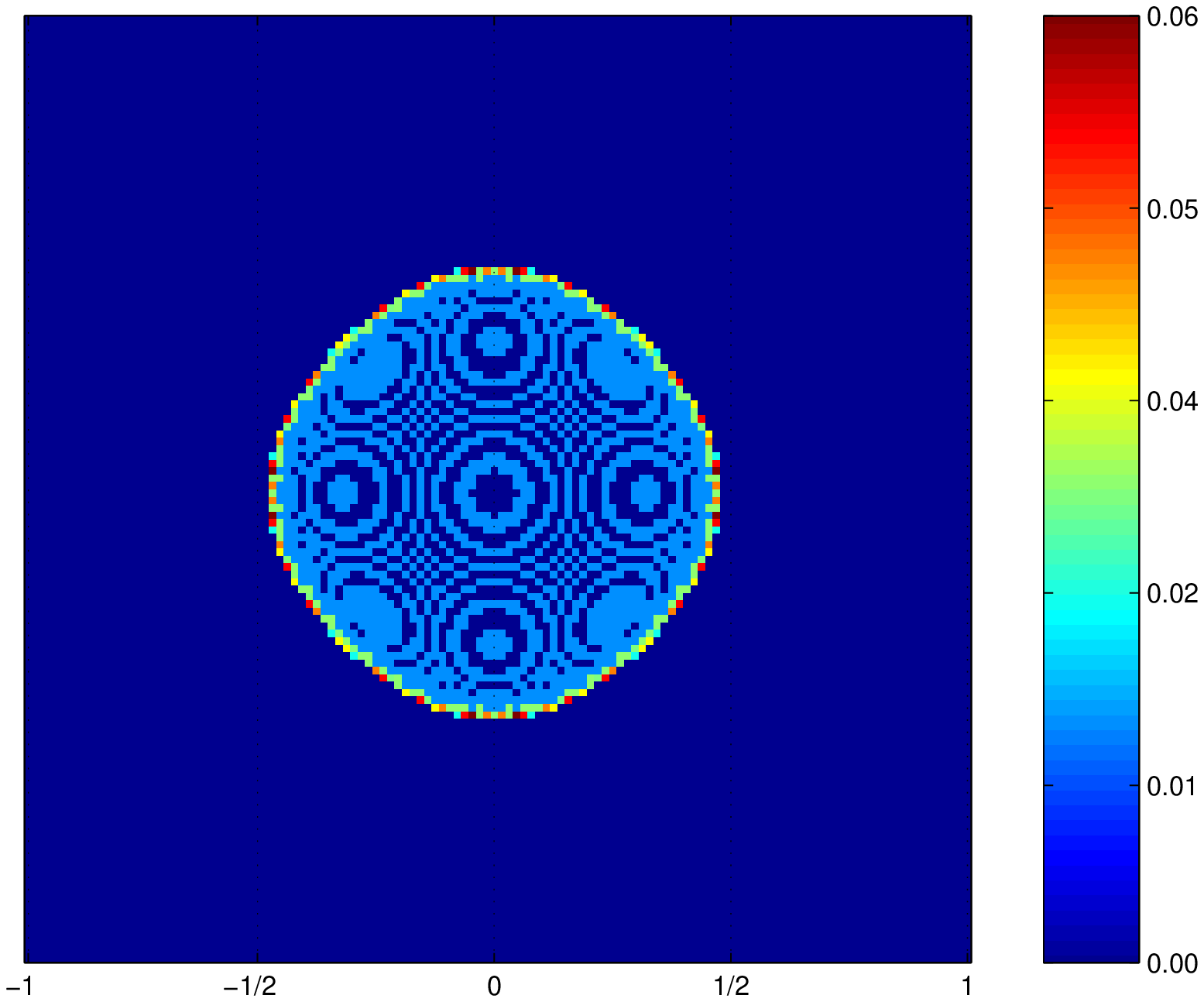}

\caption{The momentum distributions $f\up + f\down$,
  and $f\up - f\down$, integrated over the $z$-direction. Here the total filling fraction is $f=0.1$
  and $P = 0.02$. Here $\vec{q} = 0$ and the momentum difference shows
  the BP depairing symmetrically along the Fermi surface.}
\label{fig:momentum_f0.05P0.02}
\end{center}
\end{figure}

\begin{figure}
\begin{center}
\includegraphics[width=0.48\textwidth]{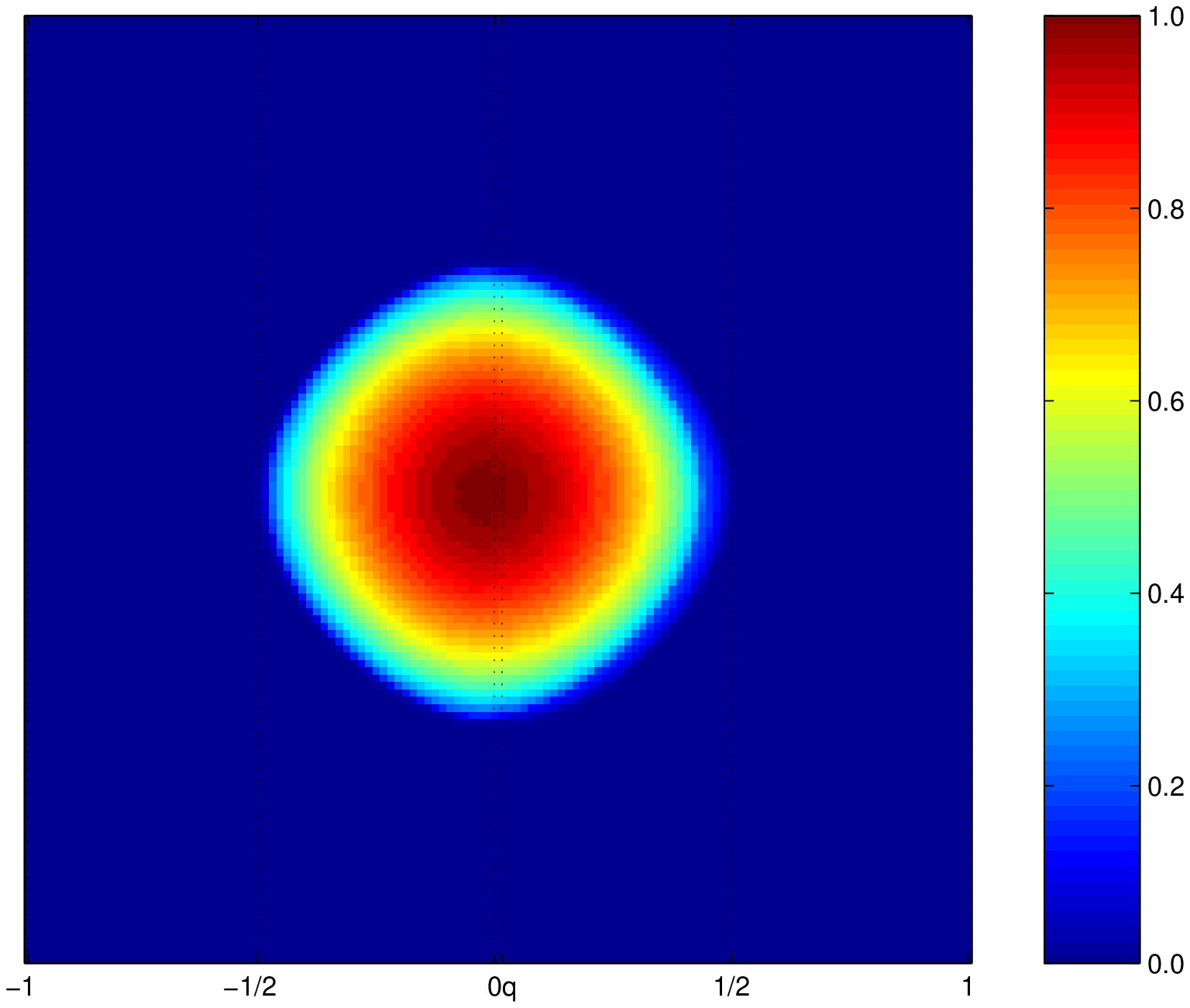}
\includegraphics[width=0.48\textwidth]{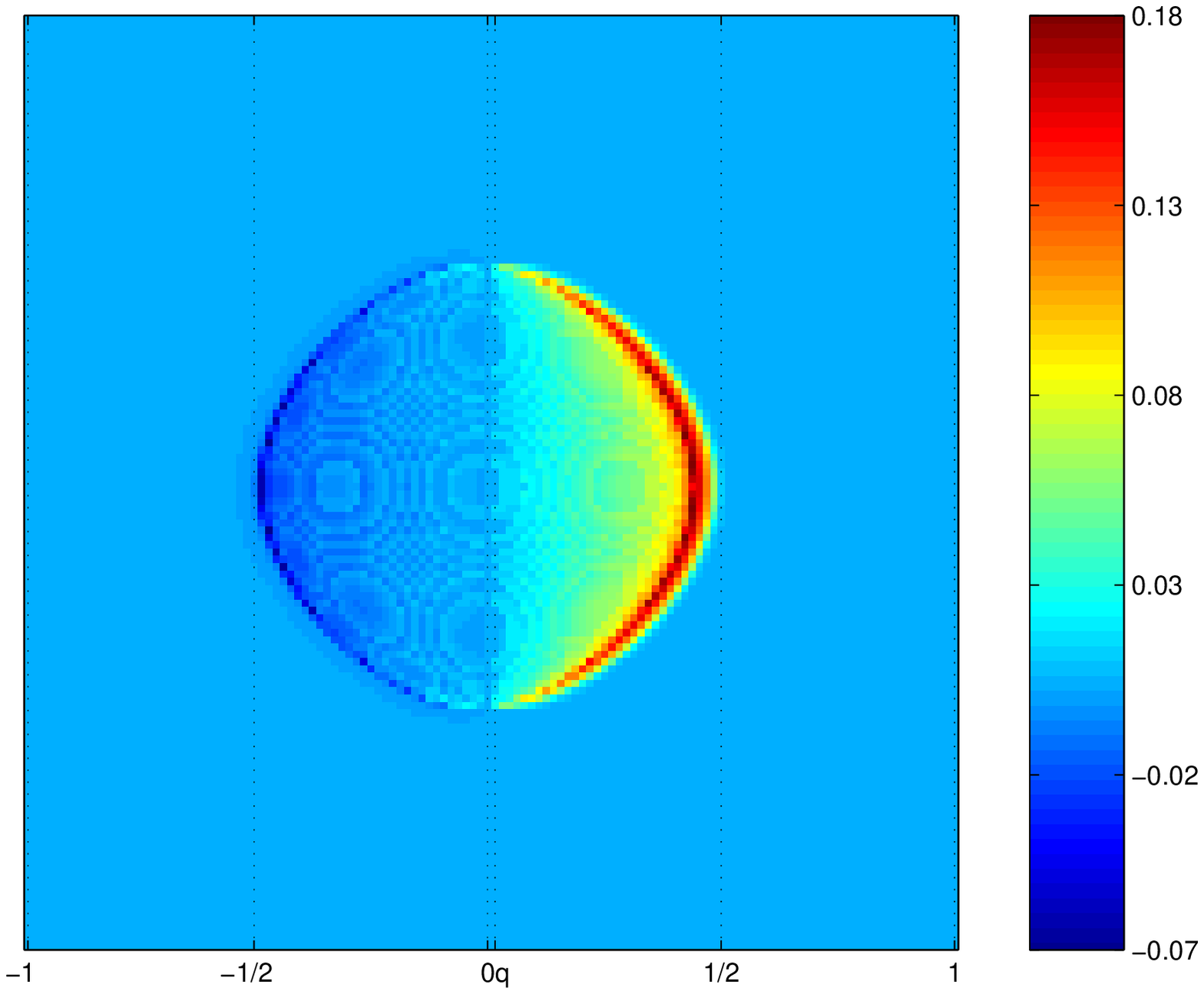}

\caption{The momentum distributions $f\up + f\down$,
  and $f\up - f\down$, integrated over the $z$-direction. Here the total filling fraction is $f=0.1$
  and $P = 0.04$. The state is of the FFLO type with a finite $q = \pi/(128d)$,
  which shows clearly in the momentum difference. The depairing region
  is similar to what has been predicted for homogenous systems, see \cite{Takada1969a}.}
\label{fig:momentum_f0.05P0.04}
\end{center}
\end{figure}

\begin{figure}
\begin{center}
\includegraphics[width=0.48\textwidth]{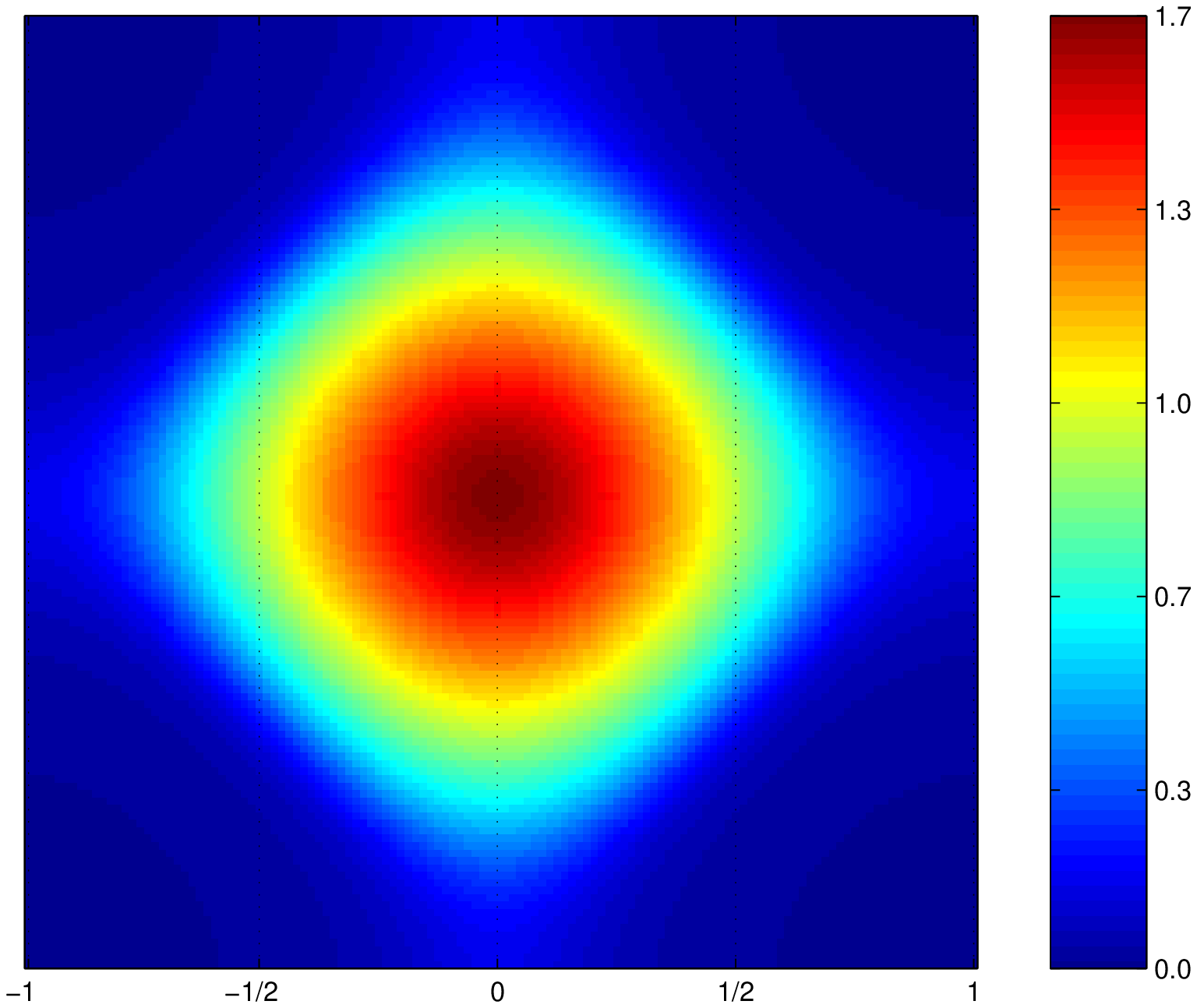}
\includegraphics[width=0.48\textwidth]{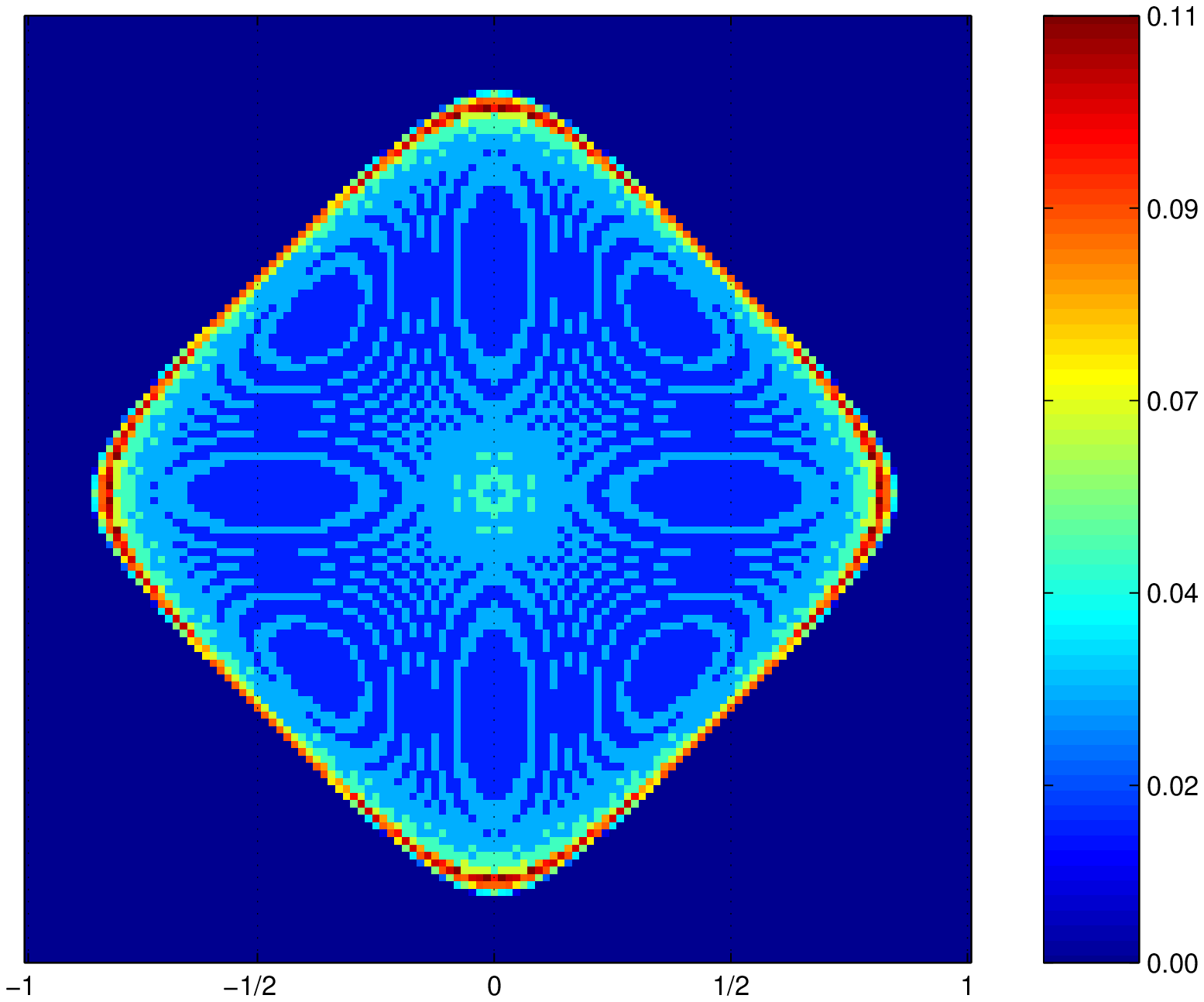}

\caption{The momentum distributions $f\up + f\down$,
  and $f\up - f\down$, integrated over the $z$-direction. Here the total filling fraction is $f=0.4$
  and $P = 0.03$. Here again $\vec{q} = 0$ and the momentum difference shows
  the BP depairing symmetrically along the Fermi surface.}
\label{fig:momentum_f0.2P0.03}
\end{center}
\end{figure}

\begin{figure}
\begin{center}
\includegraphics[width=0.48\textwidth]{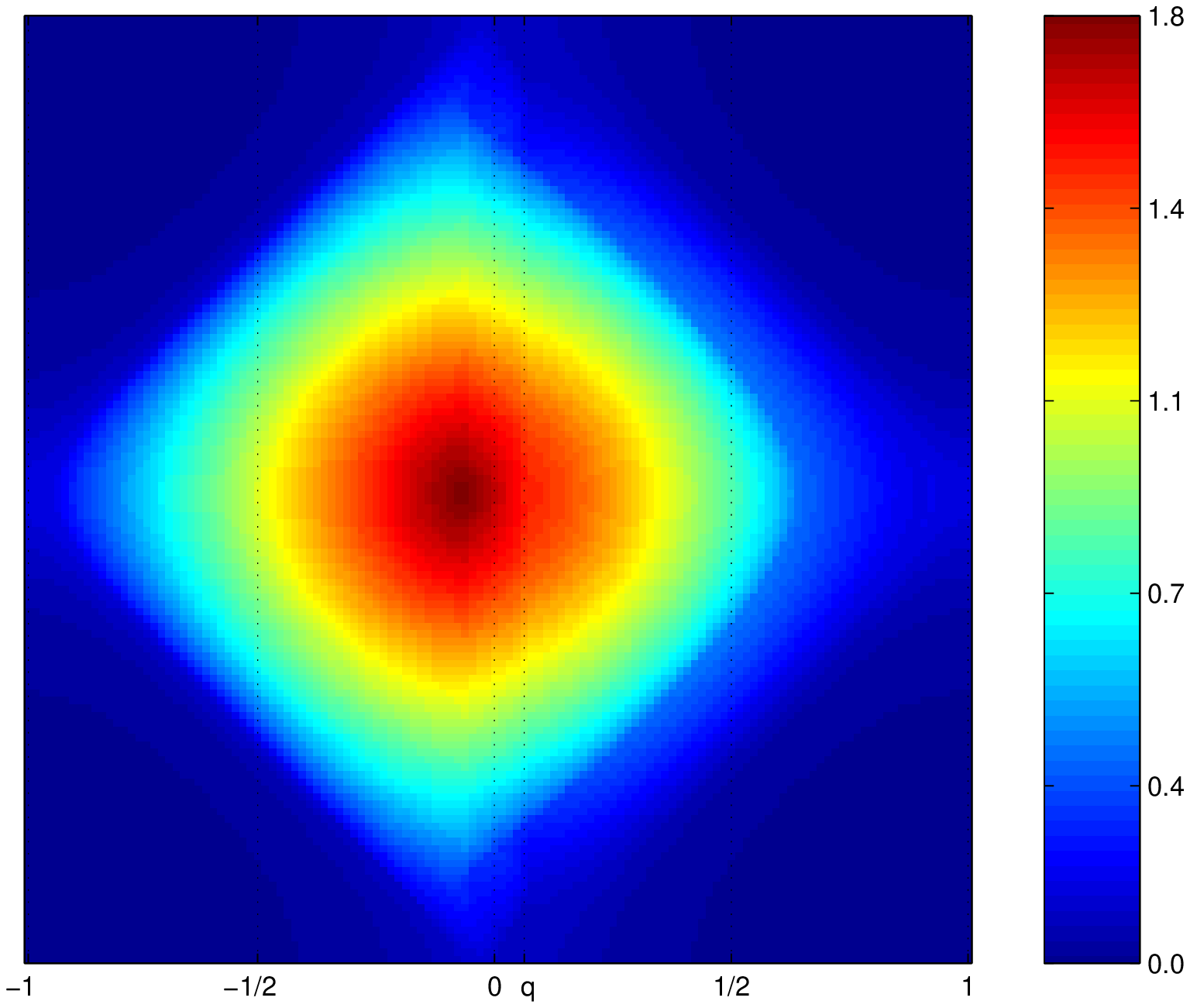}
\includegraphics[width=0.48\textwidth]{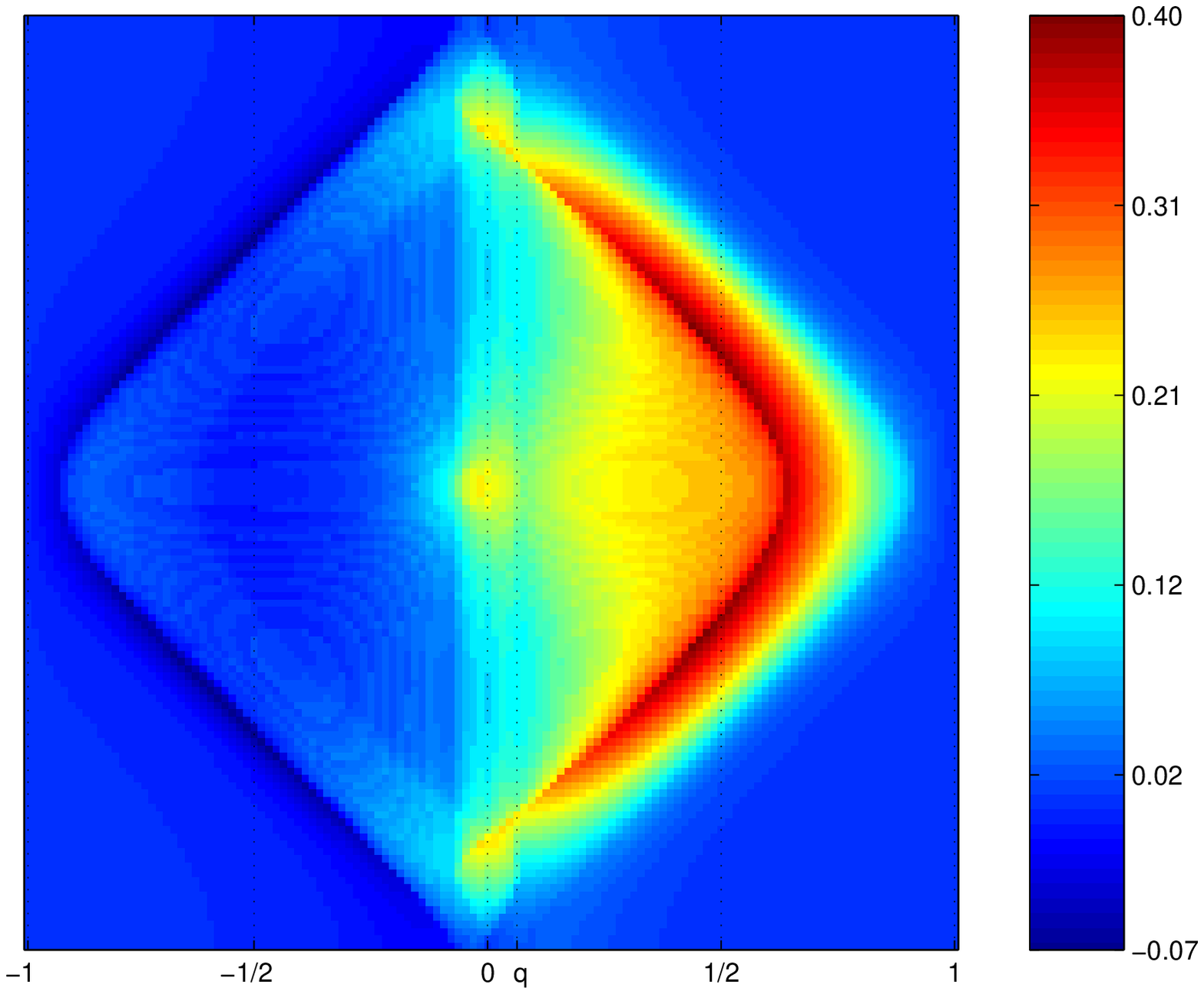}

\caption{The momentum distributions $f\up + f\down$,
  and $f\up - f\down$, integrated over the $z$-direction. Here the total filling fraction is $f=0.4$
  and $P = 0.15$. The state is FFLO with $q = 2\pi /(64d)$ in the $x$-direction.}
\label{fig:momentum_f0.2P0.15}
\end{center}
\end{figure}

\begin{figure}
\begin{center}
\includegraphics[width=0.48\textwidth]{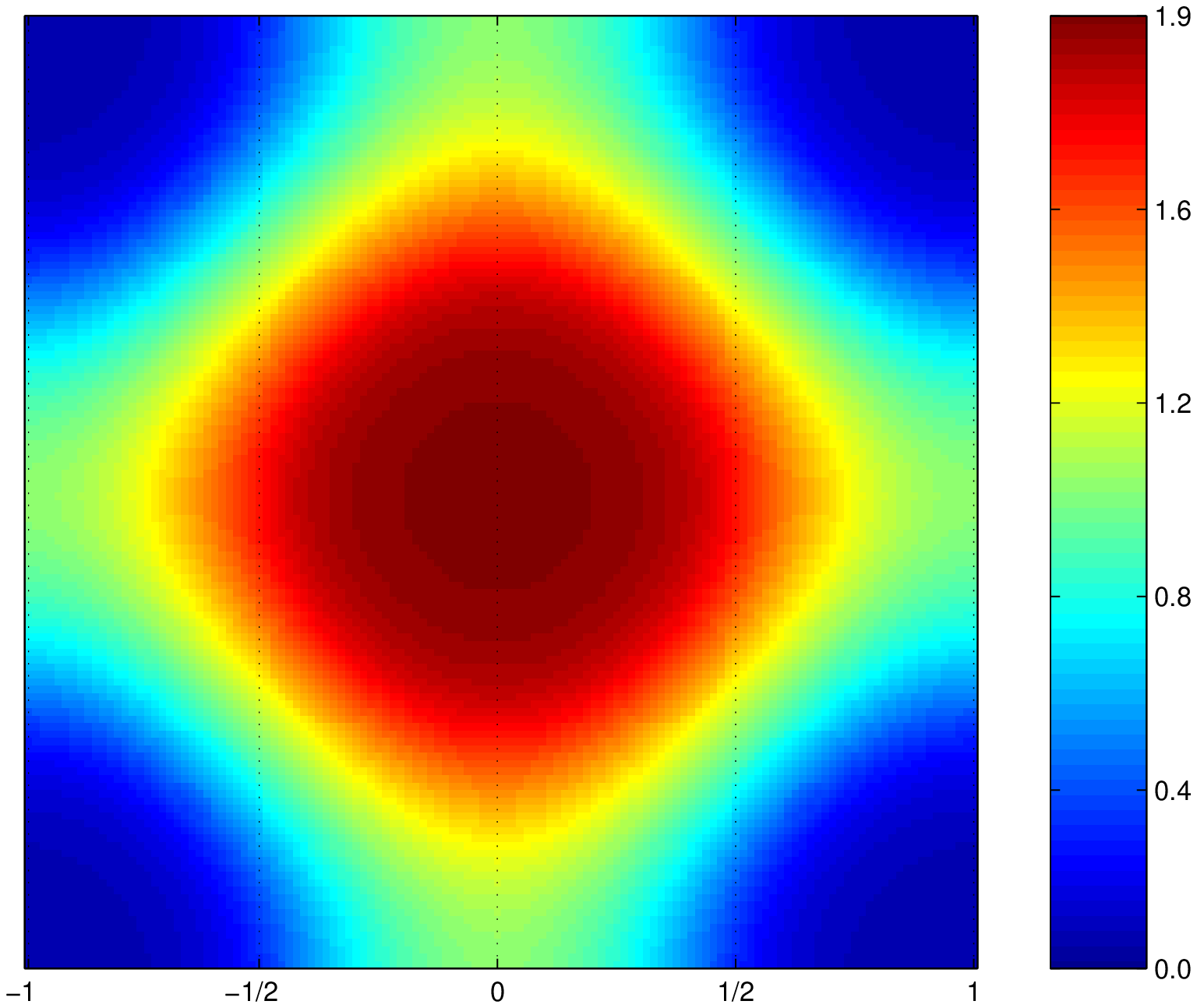}
\includegraphics[width=0.48\textwidth]{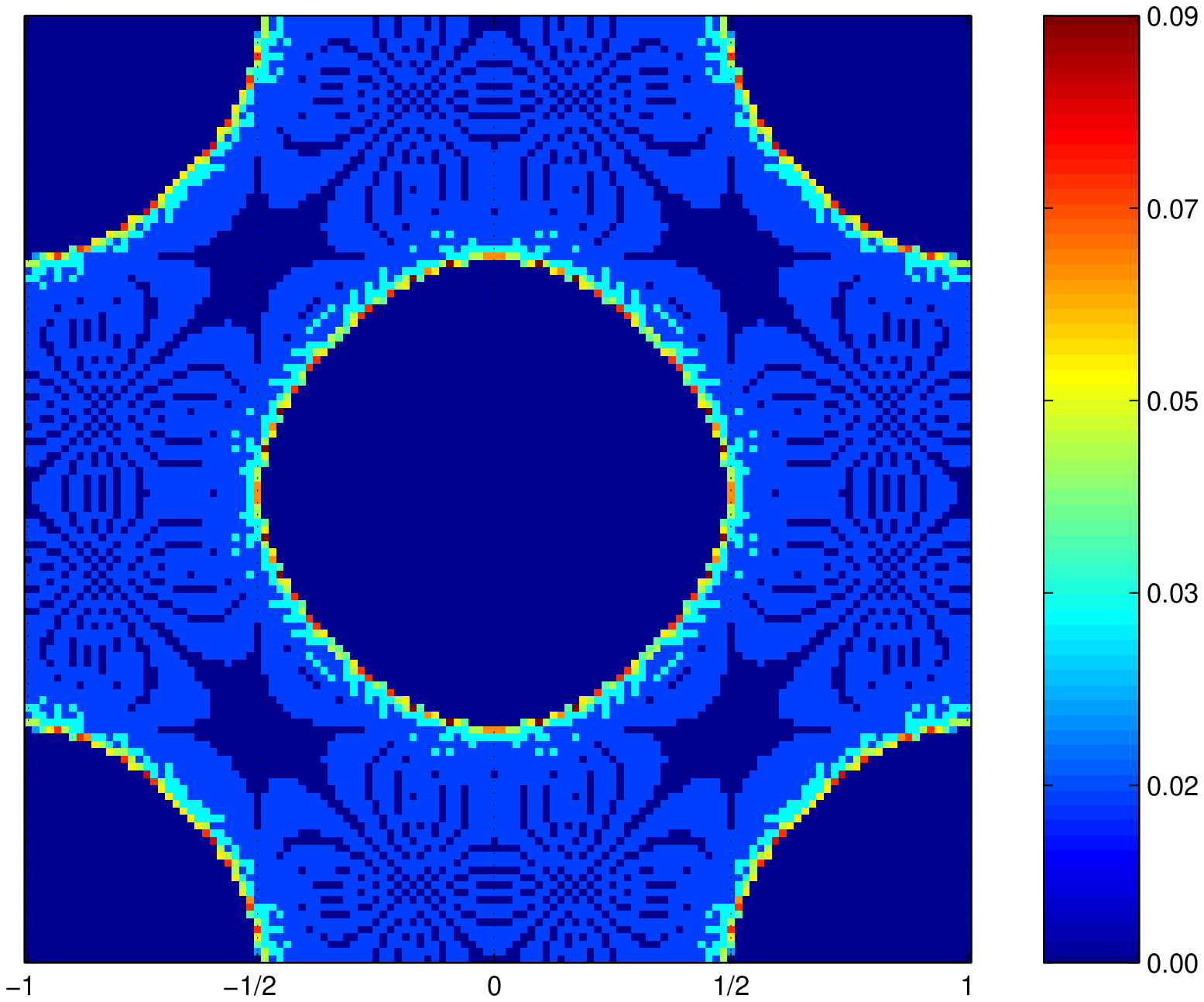}

\caption{The momentum distributions $f\up + f\down$,
  and $f\up - f\down$, integrated over the $z$-direction. Here the total filling fraction is $f=1.0$
  and $P = 0.01$. Here the state is BP.}
\label{fig:momentum_f0.5P0.01}
\end{center}
\end{figure}

\begin{figure}
\begin{center}
\includegraphics[width=0.48\textwidth]{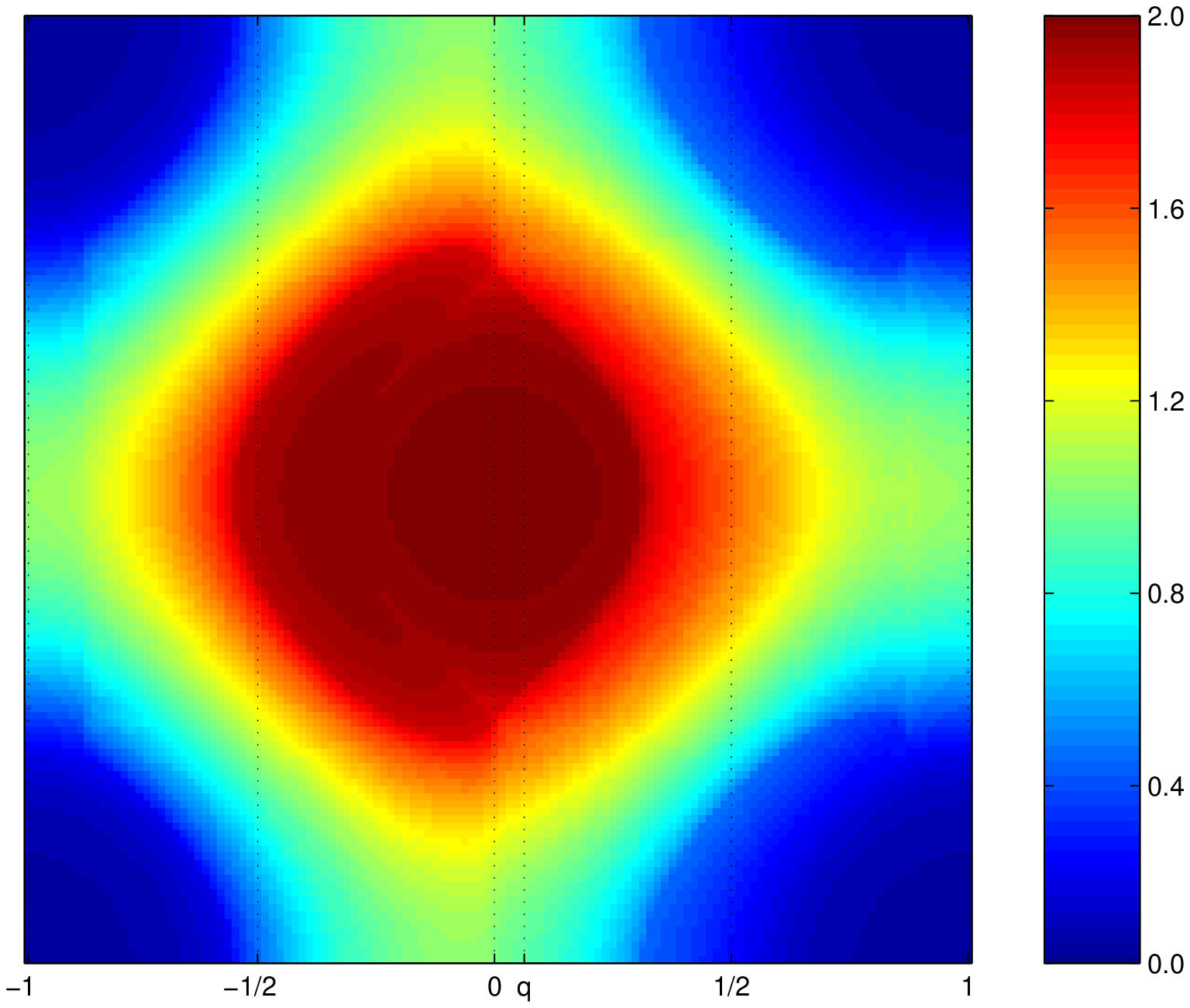}
\includegraphics[width=0.48\textwidth]{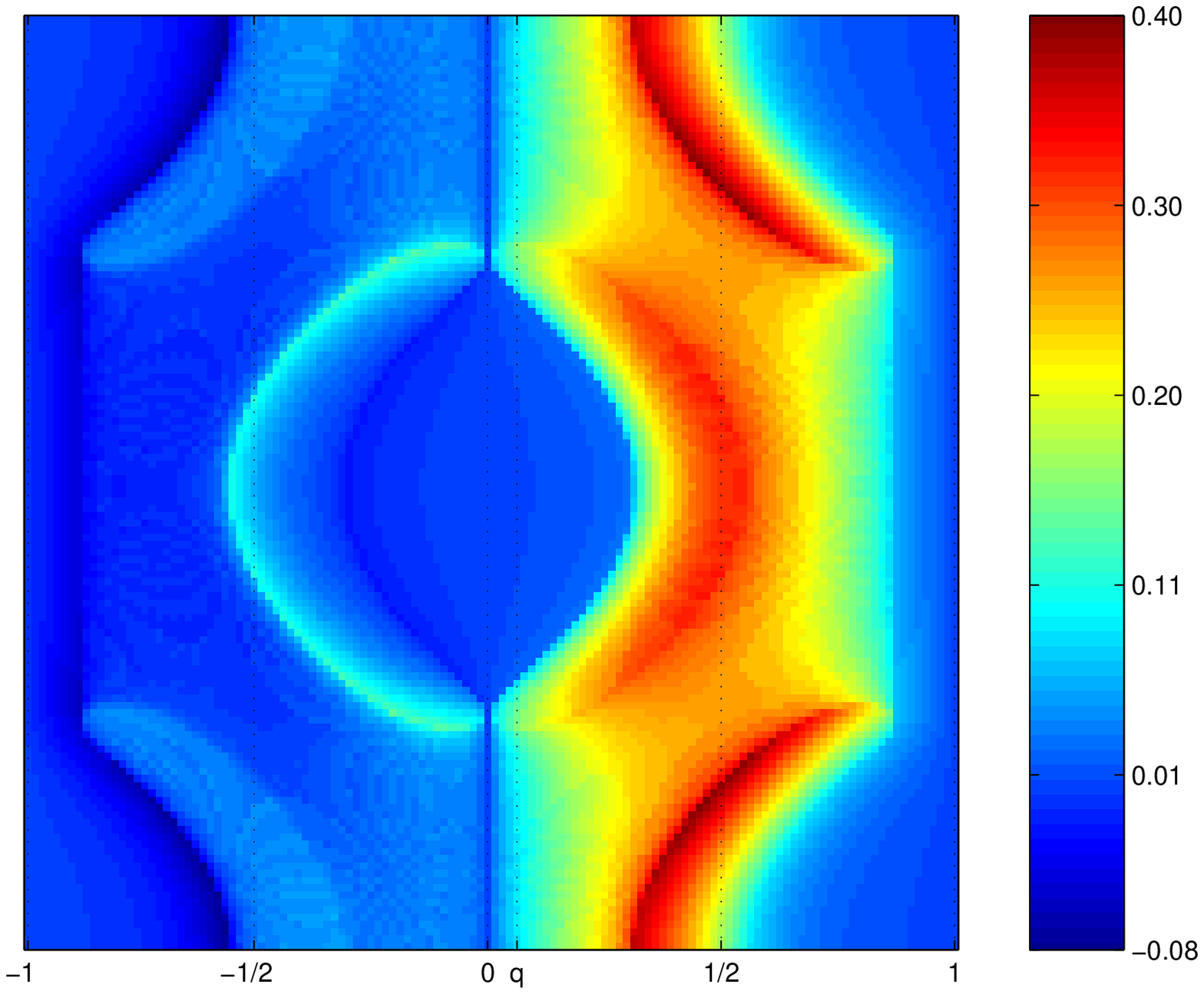}

\caption{The momentum distributions $f\up + f\down$,
  and $f\up - f\down$, integrated over the $z$-direction. Here the total filling fraction is $f=1.0$
  and $P = 0.08$. Here the state is FFLO with $q = 2\pi /(64d)$ in the $x$-direction.}
\label{fig:momentum_f0.5P0.08}
\end{center}
\end{figure}

Because the Cooper pairs each carry momentum \(2\vec{q}\), it would
seem that the system has total momentum. However,  it has been shown for homogenous systems that the
net momentum is canceled by momentum distributions of the individual
components canceling the
effect of the momentum $q$ \cite{Takada1969a}. This is consistent with figures
\ref{fig:momentum_f0.05P0.02} - \ref{fig:momentum_f0.5P0.08} where the
total momentum distribution $n_k$ ($=f\up +f\down$) is
biased to the direction opposite to where $q$ is located. We have also
numerically checked that the net momentum is zero.

Figures \ref{fig:momentum_f0.05P0.02} - \ref{fig:momentum_f0.5P0.08}
show the momentum distributions in different states, integrated over
the $k_z$. The background scattering length is $-1000\ a_0$ in each
figure. The figures show clearly the effect of filling fraction, with
low filling fractions giving a spherical Fermi surface, but higher
filling fractions showing deformations caused by the lattice. The BP state is visible in the
difference of the momentum distributions, $n\up[\vec{k}] -
n\down[\vec{k}]$, as a symmetric depairing region, whereas the FFLO
phase has asymmetric depairing. Note that the vacant region
in the momentum distribution in the BP phase is present, but not
visible due to the column integration and also due to the small polarization. 

The FFLO state has been suggested to be observable in the correlations
in the atom shot noise\cite{Yang2005a}, however, as figures
\ref{fig:momentum_f0.05P0.02} - \ref{fig:momentum_f0.5P0.08} show, the
FFLO state is already reflected in the momentum distributions.

\section{Conclusions}
We have shown that, in the case of interacting fermionic atoms trapped
in optical lattices, a stable FFLO state can be found. The stability
of various non-BCS superfluids such as BP state/Sarma state has been a
intriguing topic of study and it is known that the stability depends
for instance on whether the particle number in the system is fixed or
not. Moreover, one has to always consider also the possibility of a
state beyond the BP/Sarma state, such as FFLO-type states associated
with a non-zero Cooper pair momentum. FFLO state has been predicted to
lower the system free energy compared to the BP state in many cases,
however, it is often predicted to occur only in some narrow parameter
window.\cite{Fulde1964a,Larkin1965b} According to the analyses we present here, in atomic gases trapped in optical
lattices, the FFLO state minimizes the relevant free energy, and does
it for a considerable range of parameters. This is influenced by the
fact that the particle numbers, not the chemical potentials, are fixed
in trapped atomic gases. Furthermore, considerable critical
polarizations can be achieved, which may be partly aided by lattice
features. For low filling fractions, e.g.\ $f=0.1$, critical polarizations are
$P_c=0.075$, $P_c=0.5$, and $P_c=0.9$ for the scattering lengths
$-1000a_0$, $-1500a_0$ and $-2000a_0$, respectively, showing strong dependence on the
interactions. Also raising the filling fraction increases the
critical polarization, for instance for $-1000a_0$ scattering length,
$P_c = 0.3$ for
$f=0.4$ (compared to $P_c=0.075$ for $f=0.1$). Interestingly, however,
further increase of $f$ does not necessarily make $P_c$ to grow, for
example $P_c = 0.3$ also for $f=1$. This dependence of the critical
polarization on the filling fraction requires further study and may be
useful in comparison of theory to experiments. Stronger interactions
may bring in additional interesting features, but then one has to
consider the validity of the single band approximation. We have also
showed that the FFLO state is clearly reflected in the directly
observable momentum distribution of the atoms. 

An important topic of further study is to evaluate the effect of the
residual harmonic trapping potential which is always associated with optical 
lattices, despite their approximate periodic translational
invariance. Issues related to potential phase separation have to be
clarified. Based on previous experiments on bosonic and fermionic atoms in optical lattices, it seems possible to limit the effect of the harmonic potential in
such a way that essential predictions of a homogeneous lattice model
can be observed. On the other hand, for strong harmonic trapping,
interesting combined lattice and phase separation effects may be seen.

\ack This work was supported by the Finnish Cultural Foundation, the
Academy of Finland, and EUROHORCs (EURYI, Academy projects
Nos. 106299, 205470, and  207083).

\section*{References}
\bibliographystyle{h-physrev}
\bibliography{bibli}

\end{document}